\pdfoutput=1

\documentclass[11pt]{article}

\usepackage{acl}
\usepackage{times}
\usepackage{latexsym}
\usepackage[T1]{fontenc}
\usepackage[utf8]{inputenc}
\usepackage{microtype}
\usepackage{inconsolata}
\usepackage{graphicx}
\usepackage{soul}
\usepackage{enumitem}
\usepackage{booktabs}
\usepackage{colortbl}
\usepackage{csquotes}
\usepackage{newunicodechar,graphicx}

\DeclareRobustCommand{\okina}{%
  \raisebox{\dimexpr\fontcharht\font`A-\height}{%
    \scalebox{0.8}{`}%
  }%
}
\newunicodechar{ʻ}{\okina}

\title{Attention to \textit{Non}-Adopters}

\author{
  Kaitlyn Zhou$^{1,2}$ 
  Kristina Gligorić$^{1,3}$ 
  Myra Cheng$^1$ 
  Michelle S. Lam$^1$ 
  Vyoma Raman$^{1,4}$ \\
  \textbf{Boluwatife Aminu}$^1$ 
  \textbf{Caeley Woo}$^1$ 
  \textbf{Michael Brockman}$^1$ 
  \textbf{Hannah Cha}$^1$ 
  \textbf{Dan Jurafsky}$^1$\\
  $^1$Stanford University, 
  $^2$Together AI,\\
  $^3$Johns Hopkins University, 
  $^4$Cornell University \\
  \texttt{katezhou@stanford.edu}
}

\begin{document}
\maketitle

\begin{abstract}
Although language model–based chat systems are increasingly used in daily life, most Americans remain non-adopters of chat-based LLMs --- as of June 2025, 66\% had never used ChatGPT \cite{sidoti_mcclain_2025_chatgpt_use}. At the same time, LLM development and evaluation rely mainly on data from adopters (e.g., logs, preference data), focusing on the needs and tasks for a limited demographic group of adopters in terms of geographic location, education, and gender. In this position paper, we argue that incorporating \textit{non-adopter} perspectives is essential for developing broadly useful and capable LLMs. We contend that relying on methods that focus primarily on adopters will risk missing a range of tasks and needs prioritized by non-adopters, entrenching inequalities in who benefits from LLMs, and creating oversights in model development and evaluation. To illustrate this claim, we conduct case studies with non-adopters and show: how non-adopter needs diverge from those of current users, how non-adopter needs point us towards novel reasoning tasks, and how to systematically integrate non-adopter needs via human-centered methods.
\end{abstract}
\section{Introduction}
Large language models (LLMs) have made tremendous progress toward supporting humans in everyday tasks. While there is a substantial population of adopters -- people who use these systems -- there remains a large population of \textit{non-adopters}. For example, as of June 2025, 66\% of Americans have never used ChatGPT, the most popular commercial LLM chat model \cite{sidoti_mcclain_2025_chatgpt_use}.\footnote{\textbf{Chat models}: We refer to conversational LLMs with standalone interfaces as chat models (e.g., ChatGPT, Claude, or Gemini), limiting the scope of our research to a technology where users are \textit{intentionally} interacting with it. This definition excludes LLMs that are embedded into other systems like auto-text completion or search summarization.} Furthermore, adoption is not random but stratified, with lower adoption rates in historically technologically underrepresented demographic groups  \citep[i.e., rural, less educated, women]{mcclain2024americans, SUAREZ2025102997}.

\begin{figure}[h]
    \centering
    \includegraphics[width=1.05\linewidth]{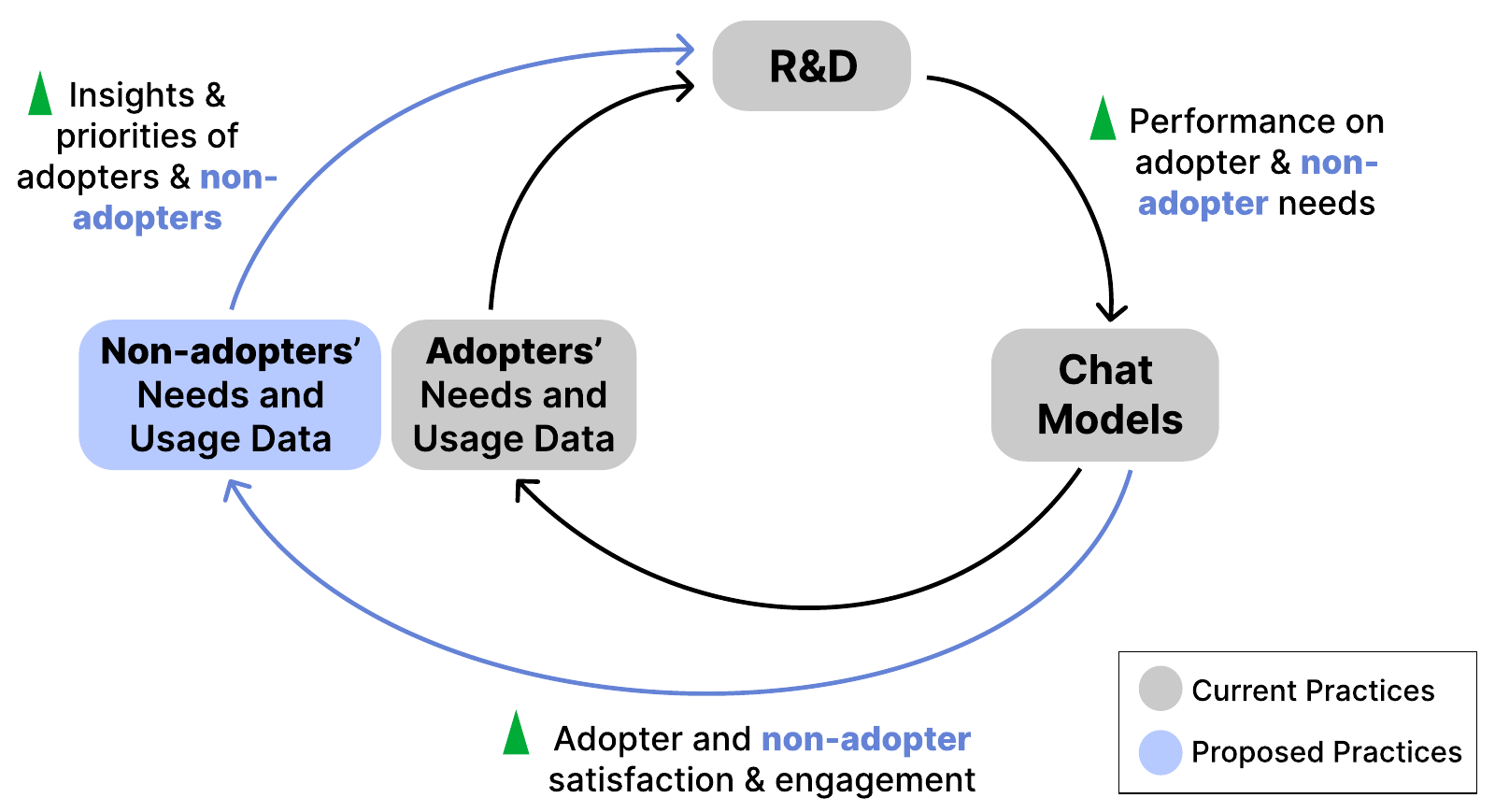}
    \caption{Proposed development cycle with the inclusion of non-adopter needs and preferences.}
    \label{fig:figure_1}
\end{figure}

At the same time, LLM developers lean heavily on interaction logs, preference data, and user studies to study and improve LLMs, including informing us of the current (\textit{and future}) use cases and failures of LLMs \cite{ding2022impact, pei2023annotator, wang2022aidreamsearchaspiration, zhao2024wildchat,tamkin2024clio, devgpt}. Although these methods provide us with convenient insights, they necessarily center on current adopters, overlooking the needs of a broader (\textit{potential}) population. 

In this position paper, we argue that \textbf{developing broadly useful chat models requires incorporating non-adopters’ needs into chat model design, evaluation, and deployment}. We take the position that current LLM development centers on the needs of current users, which represents a narrow demographic group and risks 1) failing to meet the needs of a broad and distinct user audience and 2) failing to advance LLM capabilities on a more diverse set of contexts and tasks. 

To illustrate our position, we conduct a set of case studies based on qualitative interviews ($n = 23$) and an online survey ($n = 230$) with non-adopters to shed light on the new attitudes, challenges, and unmet needs of non-adopters. 
Our studies reveal two main findings:
(1) rather than being driven by resistance to AI, many non-adopters articulate a tension between their desire to use chat models and the specific difficulties they face in learning how to use new technologies, and (2) tasks prioritized by non-adopters are distinct from those of adopters and are under-represented in chat model evaluations (e.g., navigating basic digital services like healthcare portals, coordinating caregiving responsibilities, accessing contextualized information etc.).

The bias of focusing on existing users has long been studied in HCI through the lens of non-use \cite{wyatt2002they, wyatt2005non, satchell2009beyond}. In NLP, current adopter-centered practices risk widening the divide between adopters and non-adopters as training datasets, benchmarks, and evaluation metrics are molded to meet the needs of current users. By making this position statement, we hope to direct attention toward the needs of non-adopters and to translate foundational principles from participatory design and inclusive technology adoption to develop a concrete research agenda for the NLP community. We present the following:

\begin{enumerate}

    \item \textbf{Position (\S ~\ref{sec:section_0})} We introduce the argument and background for systemically engaging non-adopters. 
    
    \item  \textbf{Why Should We Care? (\S \ref{sec:section_1})} We present two case studies ($n=23, n = 230$) to illustrate why it's important for researchers to consider non-adopters, highlighting that non-adopters are not all active resisters, and that they represent a demographically distinct user audience.

    \item \textbf{Non-Adopters Point Us Towards Novel Tasks (\S \ref{sec:section_2})} We present novel tasks that emerge from non-adopter needs (e.g., navigating basic digital services, managing domestic labor, planning physical tasks), and engage with the NLP literature to illustrate how they differ from current tasks and evaluations. 
    
    \item \textbf{Practical guidelines (\S \ref{sec:section_3})}  We identify existing practices that may inadvertently exclude non-adopter perspectives (e.g., 5\% of U.S. online crowdworkers have never used chat models, a stark contrast to the 66\% observed in the general U.S. population \S \ref{sec:upweighting_non_adopters}) and provide recommendations for uplifting non-adopter needs (e.g., re-balancing data annotation and interaction logs, participatory design for developing evaluations, and non-adopter-centered task ideation). 
\end{enumerate}
\section{Position: Attention to Non-Adopters} 
\label{sec:section_0}
Our position is motivated by two key concerns in the adoption and development of chat models: the inequity that arises when language technologies are adopted by only a portion of the population, and the narrow scope of model evaluation when it centers solely on the preferences and tasks of current adopters.

\paragraph{Concern \#1: Inequities In Model Access}
The current development of chat models is at risk of creating a new \textit{digital divide} where some have access to reap the benefits of chat models while others are shut out \cite{van2003digital}. Despite the growing financial cost of using language models \cite{openai_chatgpt_pricing, anthropic_claude_pricing, google_ai_plans}, \textit{``access,''} here, is primarily not a physical constraint, but rather a constraint based on the usability and applicability of the technologies. We argue that the current de facto focus on early adopters in data collection and evaluation is not neutral, but rather actively embeds the preferences of a tech-savvy demographic into the foundation of models, which can have far-reaching consequences that may be unknown to us today. A historical parallel comes from automotive safety, where seatbelts were primarily tested on male bodies, and women today continue to experience higher injury and fatality rates as a consequence of this exclusionary design practice \cite{weiss2001fetal, bose2011vulnerability}. In the context of chat models, we speculate that adopter-centric design risks encoding dominant narratives of how technology ``should'' be used, perpetuating a cycle where non-adopter needs remain unsupported and their future participation continues to decline \cite{barocas2020fairness, blodgett2020language}.

\paragraph{Concern \#2: Incomplete Model Evaluations} 
The frequent usage of data from early adopters risks exerting an out-sized and potentially narrow influence on the way chat models are developed. Section \ref{sec:section_1} provides initial evidence of this already happening. User data through preferences, interaction logs, and user studies are commonly used to understand user needs and influence the future direction and development of chat models \cite{jiang2024investigating, chen2024retrospective, liu2025user}. However, centering development on existing adopters risks overlooking a diverse and distinct group, and may ultimately lead to an unrepresentative sampling of interactions that doesn't span the full range of real-world human tasks. Furthermore, while chat models developed on current user data \textit{may} display generalist capabilities, if non-adopter needs remain peripheral to chat model development, such opportunistic or incidental uses will lack the systematic design, evaluation, and iteration needed for robust performance. 

\paragraph{Proposal} 
We propose that paying attention to non-adopters can address both the burgeoning digital divide and the incompleteness of model evaluations. First, preventing a chat model digital divide is an interdisciplinary challenge requiring both interface-level interventions and foundational changes to model development. Our position, given that we are NLP researchers, will focus on recommendations for NLP practitioners, such as methods, artifacts, and values held in the community. Guided by inclusive design principles \cite{clarkson2013inclusive}, we advocate for the systematic integration of non-adopters’ needs at the \textit{outset} of chat model development --- shaping datasets, tasks, and evaluations from the beginning. Second, in addressing incomplete model evaluations, the research community has an opportunity to broaden the scope of model evaluation by incorporating non-adopter needs and rigorously testing model performance in new, underrepresented contexts. When future researchers claim \textit{state-of-the-art} capabilities, non-adopter tasks should not be seen as optional, but considered core capabilities that must be robustly evaluated. The integration of non-adopter needs will also trigger additional accountability and evaluation from model providers, holding them accountable to non-adopter use cases~\cite{shen2021everyday,lam2022endUserAudits,deng2023userEngagedAuditing}.

\paragraph{The History of Studying Non-Use}
There is a significant literature on the importance and reasons for technological non-use. Researchers have highlighted the distinction between voluntary and involuntary non-use \cite{wyatt2002they} and why non-use reasons should be considered seriously \cite{satchell2009beyond}. In arguing for paying attention to chat model non-adopters, we extend the influential work of \citet{wyatt2005non} who argues: \textit{``it is essential to consider the role of non-users in the development of large technical systems... rather than focusing only on the changing relationships between system builders and users.''} Many non-adopters will have legitimate reasons to resist adopting or to stop using chat models; however, as research practitioners who have the power to design and shape model capabilities, we urge the community to develop technologies where non-use is a \textit{choice}, rather than an inevitable circumstance.

\paragraph{Ethical Concerns}
Participatory and co-design methods can accommodate non-adopter needs, but when applied uncritically, they can reproduce or amplify existing power asymmetries \cite{harrington2019deconstructing} and become exploitive \cite{sloane2022, Cooper_2024}. Since non-adopters tend to be among historically underrepresented demographic groups \cite{mcclain2024americans}, these methods must be implemented carefully to avoid widening the digital divide. Critical use, here, requires both inclusion and safeguarding against extractive practices.

\section{Why Should We Care?} 
\label{sec:section_1}
\textit{Why is it important for chat model researchers and developers to consider non-adopters?} In this section, we illustrate that non-adopters are \textit{potential users} of chat models, but because their \textit{demographics} and \textit{needs} differ from those of current adopters, they need to be explicitly accounted for.

\paragraph{Case Studies Overview}
To test our main position, we conducted \textbf{Case Study 1}: user interviews with non-adopters ($n=23$, ages 20-67, across nine different U.S. states) and \textbf{Case Study 2}: an online survey via Prolific ($n=230$, $136$ non-adopters and $94$ adopters) to illustrate how this population is distinct and can provide new perspectives on chat model development.\footnote{\textbf{Non-adopters}: Defined here as participants who rarely (i.e., once every couple of months) or never use chat models versus \textbf{adopters} who use chat models a few times a week or every day. This separates participants who have nearly no exposure to chat models from those who have integrated this tool into their everyday lives.}  These studies were IRB approved, and all participants were paid at least \$15 USD an hour. For reproducibility, see \S \ref{sec:appendix_methods_details} for recruitment details, interview scripts, participant demographics, and survey.

\paragraph{Generalizability of Qualitative Studies}
Unlike quantitative approaches, our use of qualitative methods is not intended to produce universal claims about the diverse and multifaceted group that is non-adopters. Rather, our aim is to build intuition and map the landscape of who non-adopters are, including how they engage with, resist, or remain ambivalent toward chat models. Our sample size ($n=23$) is comparable to prior qualitative studies in NLP and HCI \cite{expectation_copilot, zhou2022deconstructing, kwon2024unveiling, taranukhin2024empowering}, which included 24, 18, 12, and 15 participants, respectively. We focus on English-speaking, U.S.-based participants to situate the study within a specific socio-technical context and enable coherent analysis. While this scope limits generalizability, it offers a foundation for future work across different languages, cultures, and regions.

\subsection{Non-Adopters as Potential Users}
First, we must recognize that non-adopters make up the \textbf{majority} of the U.S. population. Work in June 2025 from the Pew Research Center finds that 66\% of Americans have never used ChatGPT and 20\% have never even heard of it \cite{sidoti_mcclain_2025_chatgpt_use}. Not only do non-adopters make up a significant population, but more importantly, our findings suggest that many non-adopters may become future users if given the right opportunities and usability. In our interviews, we find that many non-adopters are not committed to avoiding chat models, but rather, are confronted with obstacles that inhibit easy adoption. For example, non-adopters describe the adoption of technology as a painful learning process that requires labor, time, and support from others.

\begin{displayquote}
\small
\textit{``I'm somewhat bad [at] technology... and no one wanted to help me...} [P4]
\end{displayquote}

\begin{displayquote}
\small
\textit{``I'm in the generation... where there was no technology when I started... I definitely have had to learn a lot along the way... it can be a challenge for me, but I mean I love it, and I hate it at the same time.''} [P16]
\end{displayquote}

Despite these learning challenges, they still express a desire to learn and adopt chat models if given the opportunity. 

\begin{displayquote}
\small
\textit{``I heard about [ChatGPT] through my younger sister and she said she was going to teach me on how to do it. Well, we've never been able to do it. I wish I knew about it.''} [P9] 
\end{displayquote}

\begin{displayquote}
\small
\textit{``I mean, that's something I would love to explore... I would, you know, love to learn more about... how to use it''} [P2]
\end{displayquote}

Many non-adopters may not use chat models today, but understanding their needs gives us insights into how to design these language models for a broader population. For example, the verbosity of chat models can be a barrier to participants who speak English as a second language. Participants might also struggle with typing and reading on the screen; enabling text-to-speech interactions in a larger range of models would be critical for this population. These are interface changes as well as model training changes, as preference tuning and generation sampling will likely need to be modified to achieve this accessibility.

\subsection{Non-Adopters as Demographically Distinct}

\begin{figure}[h]
    \centering
    \includegraphics[width=\linewidth]{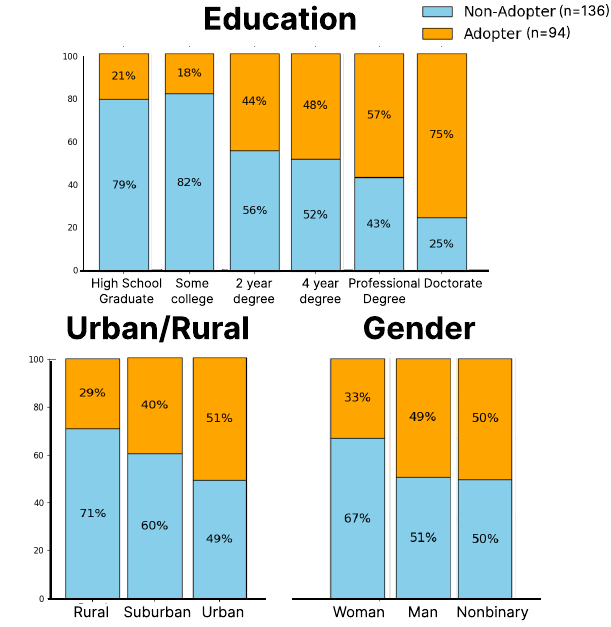}
    \caption{Education, location, and gender demographics of non-adopter and adopters. Non-adopters tend to have less than a 2-year degree, reside in rural or suburban areas, and be women. Adopters tend to have at least a 4-year degree, reside in urban areas, and are more likely to be male. *\textit{"Less than high school"} and \textit{"Prefer not to respond"} had 1 response or fewer and were omitted.}
    \label{fig:demographics}
\end{figure}
Paying attention to non-adopters allows us to design more \textbf{equitable} language technologies. The field of artificial intelligence has a long history of introducing systemic, allocational, and representation harms that impact historically underrepresented technology users \cite{bolukbasi2016man, schiebinger2014scientific, caliskan2017semantics, sheng2019woman,sap2019risk, blodgett2017racial, zhou2022richer, santurkar2023whose, blodgett2016demographic, jurgens-etal-2017-incorporating, koenecke2020racial, ogunremi2023decolonizing}. Adoption disparities for chat models intersect with model-level biases to reinforce the exclusion of groups already marginalized in technology use and design. 

Prior studies have documented demographic gaps in digital technology adoption, and the usage of chat models appears to follow similar patterns \cite{mcclain2024americans, SUAREZ2025102997}. Our survey ($n=230$) confirms these trends: current chat model adopters tend to be highly educated, urban, and predominantly male. We compared adoption rates across demographic groups using a two-proportion $z$-test, where we test for the rate of adoption. We find that participants without two-year degrees are nearly three times less likely to be adopters compared to those with professional degrees, $z$ = -3.96, $p$ < 0.001, or doctorates, $z$ = -3.99, $p$ < 0.001; participants who live in rural areas are 40\% less likely to be adopters than those living in urban area, $z$ = -2.16, $p$ < 0.05, and women are 30\% less likely to be adopters compared to men, $z$ = -2.71, $p$ < 0.01, Figure \ref{fig:demographics}.\footnote{Differences are less pronounced among demographics like race, age, and income, Figure \ref{fig:demographics_appendix}.} 

These findings highlight that the disparities between chat model adoption are not arbitrary and can be partially explained by demographic characteristics known to be underrepresented in technology access \cite{afzal2023digitaldivide, elena-bucea2021digitaldivide}. Here, we stress the urgency to pay attention to current non-adopters who represent a historically underrepresented group of technology users and mitigate the allocation harms resulting from chat models. Hence, although chat models and generative AI more broadly are touted to reduce equity gaps \cite{FU2025105347, pierson2025usinglargelanguagemodels, gabriel2024generative, nixon2024catalyzingequitystemteams}, this is not possible without directly considering and accommodating their non-adopters.
\section{Non-Adopters Reveal New Task Priorities}
\label{sec:section_2}
Non-adopters are not only demographically distinct, but they shed light on (1) real-world tasks that are missing in chat model evaluations and (2) ways that current instantiations of classic tasks like question-answering and technology navigation may be misaligned to the needs of a broader population. 

\subsection{Currently Prioritized Tasks} 
Our online survey (Figure \ref{fig:importance1}) highlights that adopters highly prioritize writing/reading, technology navigation, and creative tasks --- common tasks in the NLP literature \cite{Hadi2023LLM, kaddour2023challenges, naveed2024LLM, shao-etal-2024-assisting, gero-etal-2022-design, NEURIPS2022_82ad13ec, fereidouni2024searchqueriestrainingsmaller, zhu2023cam, chen-ding-2023-probing, tian-etal-2024-macgyver,gomez-rodriguez-williams-2023-confederacy}. These findings suggest a potential pattern of self-beneficial technology development known as ``me-search''~\cite{bradley2011me}, which occurs when researchers prioritize the needs of those who are similar to them, in this case, technology developers and current LLM adopters. Even with good intentions, the resulting technologies may end up primarily benefiting those already advantaged, leading to a rich-get-richer effect. Given the known socioeconomic, racial, and gender disparities among computer scientists \cite{pearl1990becoming, lunn2021exploration}, this trend can further exacerbate existing inequalities in LLM benefits. 

\subsection{Under-prioritized Tasks}
In contrast, over half of the non-adopters we interviewed mentioned tasks related to domestic labor. We compared task priorities across adopters and non-adopters using a two-proportion $z$-test, testing whether participants selected each task as one of their top three most important tasks. Non-adopter survey participants were over four times more likely to choose physical tasks as an important task, $z$ = 6.13, $p$ < 0.001, and 60\% more likely to choose caregiving as an important task compared to adopters, $z$ = 2.35, $p$ < 0.05. Examples include caring for younger and older children, facilitating family conversations, or supporting a loved one with physical disabilities. Many of these tasks have been recognized in the economics literature to be historically \cite{sayer2005gender} and currently \cite{petts2021gendered} performed by minoritized groups. Our interview data also reveals that these \textbf{tasks are not just physical, but also include the invisible cognitive labor known as the mental load} \cite{dean2022mental}:

\begin{figure}[h]
    \centering
    \includegraphics[width=\linewidth]{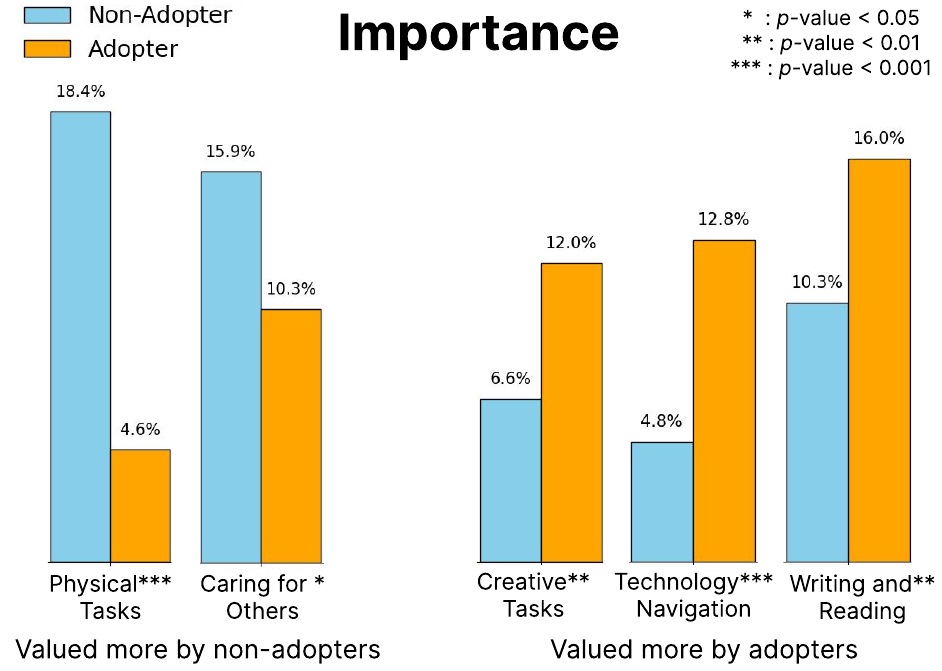}
    \caption{Task importance rankings between adopters and non-adopters with the biggest differences (denoted by \% of responders). Non-adopters tended to prioritize physical tasks and caring for others compared to adopters. Adopters tended to prioritize creative tasks, technological navigation, and writing/reading tasks.}
    \label{fig:importance1}
\end{figure}

\begin{displayquote}
\small
\textit{``[As] the mom you negotiate and [you] make sure everybody's happy... that's your main focus in life''} [P20].
\end{displayquote}

\begin{displayquote}
\small
\textit{``I always struggle because you have to coordinate with the different people, with the teacher, the decorator, make sure everybody is on the same page... keep it within budget''} [P8].
\end{displayquote}

\begin{displayquote}
\small
\textit{``[Scheduling] is a huge nightmare... If [a chat model] could know all the schedules... plug that in, and then it would tell you... `This is the best time to see... a possible mixed group of these 2 grades of students' --- that would be a huge headache reliever.''} [P6]
\end{displayquote}

Physical and domestic tasks are rare in the NLP literature and when they appear, they often prioritize technological innovation and are less situated in addressing the real-world needs or requests of specific user groups \cite{harashima-etal-2016-large, sato-etal-2016-japanese, li-etal-2022-share, morales-garzon-etal-2021-semantic, budzianowski2018multiwoz, liao2019deep, inaba2022collection}. For example, prior work on developing LLM as a scheduler, a need articulated by participant \#6, is in the context of systems optimization rather than scheduling tasks for people and their contexts \cite{pl_schedulung}. Paying attention to non-adopter needs allows us to see the importance of cognitive labor for physical and domestic tasks and recognize their under-prioritization in language model evaluations. 

\subsection{Misaligned Tasks}
Listening to non-adopters also reveals how existing tasks in chat model evaluation practically differ from what non-adopters need. The interviewees named trying to obtain high-quality information on complex topics as a key pain point. Often, they seek \textit{non-trivial} information that is specific to their unique circumstances. Our case study participants struggled particularly with insurance policies, immigration, and personal finance, and they often need personalized information unique to their identity or circumstance.

\begin{displayquote}
\small
\textit{``As a Black woman...being made aware like what to do in certain situations...diseases that are [prevalent] in like the African American population and how we [can] reduce those type of things from like occurring''} [P5].
\end{displayquote}

\begin{displayquote}
\small
\textit{``I can't even find anyone who really works with people who get pensions...I wish I could get a consultation, someone who understood where we're at''} [P16].
\end{displayquote}

Additionally, non-adopters articulate technology barriers infrequently found in NLP literature, e.g., difficulty typing, submitting documents online, or navigating health care websites. These interaction challenges are ubiquitous and highly limiting, and unaddressed needs could result in users abandoning their tasks entirely. 

\begin{displayquote}
\small
``\textit{I've been on [this] BSN program [for the past 5] years because I was not really comfortable with the computer and everything. I quit. I didn't finish''} [P1].
\end{displayquote}

\begin{displayquote}
\small
\textit{``I've been here for almost 2 years and I'm still having trouble finding a dentist... And they have all this online scheduling... their websites are horrible and don't work''} [P12].
\end{displayquote}

While tasks in information retrieval (IR) and question answering (QA) are widely studied by NLP researchers, they typically focus on factoid questions such as questions from reading passages \cite{rajpurkar-etal-2016-squad, joshi2017triviaqa}, standardized tests \cite{hendrycks2020measuring}, and mimicking historical search queries \cite{kwiatkowski2019natural, bajaj2016ms, he2017dureader}. Recent research on LLMs as agents continues to follow the pattern of retrieving information through a series of steps, querying various dense but non-contextualized materials \cite{du2025deepresearch, wan2025deepresearcharenaexamllms}. Paying attention to non-adopters highlights the need to evaluate models on non-trivial, contextualized information seeking tasks. One example of this user-centered contextualized work in NLP is from \citet{taranukhin2024empowering} who grounds their research in a concrete application --- supporting Canadian air travel passengers --- and consistently centering the needs of a clearly defined user group.

Similarly, numerous tasks in the LLM literature are dedicated to supporting technology interaction for those who are highly technical. Most salient is code generation \cite{muennighoff2024octopackinstructiontuningcode, chen2021evaluating, li2022competition, svyatkovskiy2020intellicode, chen2022codet, lai2023ds}, but this also includes plotting \cite{bubeck2023sparksartificialgeneralintelligence, yang2024matplotagent} and computer science research more broadly \cite{desai2023using, xiao2023supporting}. Meanwhile, research on web agents has focused on domains like shopping \cite{NEURIPS2022_82ad13ec, fereidouni2024searchqueriestrainingsmaller}, dining \cite{zhu2023cam}, and travel \cite{sun2022meta}. Interviews with non-adopters highlight their need for support in navigating technology—similar to technologists, though oriented toward a simpler set of tasks.

\subsection{The Importance of Non-Adopters}
Our user interviews with non-adopters give a glimpse into the complexities that go into under-prioritized tasks and reveal opportunities to develop new types of chat model evaluations. Paying attention to non-adopters' tasks opens up new challenges: \textit{Could language models support non-traditional forms of reasoning? Could language models help with the planning of physical and care-taking tasks? What are the safety risks involved?} Our position is that we must give attention to non-adopters, and we urge the NLP community to recognize the importance of asking and prioritizing the needs surfaced by non-adopters. 

 \section{Practical Guidelines: How Could We Incorporate Non-Adopter Needs?}
 \label{sec:section_3}

Natural language processing research has been fundamental to the rapid development and advancement of chat models. Here, we outline specific practices in data annotation, benchmark design, and task ideation that may be inadvertently overlooking non-adopters and how alternative practices can be used. Exclusion of non-adopters mirrors psychology’s reliance on western, educated, industrialized, rich, and democratic (WEIRD) populations, where narrow participant pools have long skewed findings and limited their generalizability \cite{henrich2010weirdest}. In an effort to counteract these effects, we propose drawing from participatory design \cite{schuler1993participatory} and co-design \cite{steen2013co, Sanders01032008}, in an effort to surface community priorities, redistribute design power, and mitigate harms when working with marginalized communities \cite{tseng2025ownershipjusthappytalk, sloane2022, harrington2019deconstructing, cruz2023equityware}. 

\subsection{Up-weighting Non-Adopters in Data} 
\label{sec:upweighting_non_adopters}
Data annotation practices may also be prioritizing the needs of chat model adopters --- further embedding adopter preferences into model training reinforces existing usage patterns. Data annotation for natural language processing has long used crowdfunding platforms like Prolific or Amazon Mechanical Turk to recruit humans to generate data, evaluate generations, gather preferences, and eliminate toxicity in language \cite{munro2010crowdsourcing, sabou2012crowdsourcing, zhao2014evaluation}. In September of 2025, we conducted a Prolific experiment with 500 U.S. participants (see \S \ref{sec: prolific_experiment} for details). We found that less than 5\% of recruited participants have never used chat models, compared to the U.S. average of 66\%, showing a huge bias of adopters as annotators \cite{sidoti_mcclain_2025_chatgpt_use}. Just as uneven annotator demographics can introduce biases into NLP datasets \cite{ding2022impact, pei2023annotator, wang2022aidreamsearchaspiration}, emphasizing adopter preferences and data will likely also introduce adopter-centered biases. 

Similarly, analysis of interaction logs has been instrumental to our understanding of chat models \cite{zhao2024wildchat,tamkin2024clio, devgpt} as it reveals the type of interactions, unexpected edge cases, and frequency of chat model usage. However, interaction logs exclusively reflect data from active and frequent users, thus skewing insights toward those with the resources and motivation to engage early and often, and lacking insights on the needs of potential users and how they might leverage LLM capabilities. 

One tested way to counteract adopter bias in data annotation and interaction log analysis can be to up-weight non-adopter preferences and labels. \citet{Gordon_2022} propose \emph{jury learning}, where researchers can designate particular groups of annotators whose inputs are weighted more heavily to reflect community values. In data labeling, non-adopter data contributions can be up-weighted to counterbalance the dominance of adopter perspectives. We applied a similar method when we recruited online participants for our own survey data, randomly filtering out chat model adopters and balancing the participant groups evenly. In interaction data, amplifying the voices of low-frequency users and engaging non-adopters through participatory or co-design methods can reveal novel and diverse model interactions.

\subsection{Realigning Benchmarks through Non-Adopter Contexts}
At the dataset level, non-adopter needs could be integrated by revisiting existing tasks with the goal of including non-adopter contexts. For example, in factoid question-answering, participatory methods could be used to revisit existing artifacts and ground questions in non-adopter needs. Examples include: understanding new ideas using familiar concepts (e.g., \textit{explaining new lingo}), explaining technical components in simple terms (e.g., \textit{What is `AirDrop'?}), and contextualizing niche scenarios (e.g., \textit{understanding pensions for retired teachers}). A concrete example stems from the work of \citet{vatsal2024can} who makes advances in the use of chat models to improve the process of prior authorizations for clinicians (i.e., advance clearance for a medical procedure). Re-contextualization here can simply extend this work to help patients understand what is covered in their insurance policies.

Tasks in technology navigation could also be realigned to meet the needs of non-adopters. Here, tasks that are essential activities like filing taxes with free software, scheduling medical appointments, or uploading files to online learning platforms could be prioritized in addition to tasks that serve advanced users. 

\subsection{Expanding Task Design}
Non-adopter needs can be leveraged to expand existing tasks, using methods like need-finding interviews or focus groups. Today, tasks that were once important in the field of natural language processing (e.g., question-answering, information retrieval, summarization, code generation, mathematical computation) still dominate in the evaluation of \textit{state-of-the-art} models. However, non-adopters express interest in a different set of tasks—such as help with the information needs associated with domestic, care-taking, and physical tasks that are rarely reflected in current evaluation benchmarks. In Table \ref{tab:sensemaking-singlecol}, we offer an example of one benchmark that operationalizes the information finding needs non-adopters expressed in \S \ref{sec:section_1} --- showing chat model development beyond factoid QA and towards complex, grounded reasoning that measurably improves human capability augmentation.

\begin{table}[h!]
\captionsetup{font=small}
\centering
\small
\renewcommand{\arraystretch}{1.2}
\begin{tabular}{p{0.95\columnwidth}}
\toprule
\textbf{Scenario}: Unexpected mobility-limiting event (e.g., sudden illness) requiring adaptation to unfamiliar domains. \\ \midrule
\textbf{User Need}: Need to research medical information, coordinate logistics, and revise daily routines. \\\midrule
\textbf{New Task}: \textit{Sensemaking} involves information synthesis, context-sensitive reasoning, and emotional attunement. \\\midrule
\textbf{Model Input}: User profile and contextual data, e.g., \textit{“I'm scheduled for surgery in a week, but until then, I need to modify my routines to minimize discomfort and decrease risk of injury. What are some things I should know?”} \\\midrule
\textbf{Model Output}: Agentic and interactive approach. The model might 1) outline a plan of action, 2) synthesize user input, 3) retrieve relevant information, and 4) develop a teaching plan. This may unfold over multiple turns where the model acts like a tutor—gradually building understanding and adapting to user questions. \\\midrule
\textbf{Evaluation Metrics}: Beyond factual accuracy, evaluation could assess user experience: \textit{Did the user feel in control of the interaction? Did they gain a better understanding of the topic? Have they learned strategies to navigate similar problems in the future?} \\
\bottomrule
\end{tabular}
\caption{New Potential Benchmark Task: Sensemaking in Unfamiliar, Time-Sensitive Contexts}
\label{tab:sensemaking-singlecol}
\end{table}

\section{Discussion and Conclusion}

\paragraph{Thinking Beyond Current Evaluation Structures}
Our work emphasizes that non-adopter tasks require thinking beyond the status quo structures of chat model training and evaluation. Non-adopter tasks are currently less structured and less objective \cite{guerdan2024indeterminate}. Training chat models on such tasks will also be challenging, as automated metrics are easier to design for structured tasks. The field requires new methods for evaluation and training, as well as greater end-user involvement, to navigate the disagreement inherent to subjective and user-centered tasks \cite{fleisig2024perspectivist, plank2022variation, gordon2022juryLearning}.

\paragraph{From Chat Model Capabilities to Human Capability Augmentation}
The integration of non-adopter tasks into chat model development warrants a paradigm shift away from evaluating chat model capabilities in isolation (\textit{``Can the model name the stages of cancer?''}) and towards evaluating \textit{human} capabilities when using chat models (\textit{``Can a user receive support through a new medical diagnosis?''}). A capabilities-augmentation evaluation paradigm extends the work of situated evaluation \cite{Arzberger2024situated}, as it argues for end-to-end support across user-defined tasks, rather than siloed subtasks defined by external researchers~\cite{mcclain2025us}. A focus on human capabilities requires a fundamental shift in the tasks we pose and the evaluation metrics by which we measure success as a field. 

In this position paper, we argue for the importance of the systematic integration of non-adopter needs into the development of chat models. We illustrate through two case studies how non-adopters differ in demographics and needs from adopters and how their unique perspectives can offer novel insights for expanding LLM evaluations. Lastly, we identify how current NLP practices can be expanded to incorporate non-adopter needs through the use of human-centered methods.  

\section{Limitations}
Due to our recruitment strategy for study participants, we exclusively interviewed individuals in the United States. Given that varied geopolitics can create different types of needs, our study does not claim to represent needs that are more prevalent in other areas. Furthermore, we conducted interviews in English, which further restricted our recruitment efforts to English speakers in the U.S. Our interview and recruitment strategies relied on internet access and the ability to use a computer or phone, so all participants have that baseline level of technical literacy despite low usage of chat models.

Although physical needs featured as an interesting finding in our study (Section \ref{sec:section_1}), we did not explicitly collect disability-related demographic information, so our evaluation of participants' disability status stems from our conversations during interviews and may be narrow in scope. We also did not interview any participants who self-identified as non-binary, which means that our findings may not reflect the unique needs of that group.

By conducting our empirical research through a single one-hour virtual interview, it is unlikely that we established a bond with participants that would have occurred with in-person or sustained synchronous contact. That means that sensitive or emotionally taxing needs may not have surfaced in our conversations, reducing the scope of our findings.

Finally, we selected OECD's \textit{The Survey of Adult Skills} to contextualize and augment our qualitative coding of interview themes. We acknowledge that this framework is premised on a theory of labor and value that may not be universally agreed upon. Nevertheless, it has proven a useful tool for categorizing the needs we identified and considering gaps in our analysis.

\section{Ethical Concerns}
Despite our focus on English-speaking U.S.-based participants, we do not advocate for a solely U.S.-centric approach to need identification and task selection. We studied the needs of one sample population that we were able to recruit. Our findings may not generalize to participants from other parts of the world or even others in the U.S., and we would not expect the needs and pain points we identify here to be the most universally salient. Rather, our work attempts to diversify perspectives around which and whose needs should be met in future LLM development, and we seek to further inspire future attempts to do so with different sample populations. 

Furthermore, the goal of our personalized demonstration of ChatGPT to interviewees was to further probe their motivations and characterize the suitability of LLMs to their needs. We did not intend to change participants' daily behaviors and habits or advertise products among non-adopters. After the demonstration, each participant was briefed about the risks of AI.

\section*{Acknowledgments}
First and foremost, we give thanks to all the participants who took part in our interviews and online surveys, whose insights and responses enabled this research.

This work was made possible with the support, feedback, and review from Sofia Kim, Jimin Mun, Jordan Taylor, Mingqian Zheng, Vasudha Varadarajan, Adam Visokay, Federico Bianchi, James Zou, Yongchan Kwon, Shang Zhu, Katelyn Mei, Dexter Crowley, Lindsay Popowski, Nava Haghighi, and Suzanne Lessard.

This work was supported by NSF grant PTA 1258785-1-QABYJ.

\bibliography{anthology, custom}

\appendix
\section{Methods Details}
\label{sec:appendix_methods_details}
Using a bottom-up approach, we begin by outlining the methods that led to our preliminary findings, which form the foundation of our position. User interviews were conducted in the spring of 2024, and an online survey was designed and released in the spring of 2025. 

\paragraph{Interview Design}
Our interview questions were designed to surface participants' needs in their work and personal lives. We draw on \textit{The Survey of Adult Skills}, developed by the Organization for Economic Co-operation and Development (OECD) which surveyed over 166,000 adults \cite{/content/publication/9789264204027-en} to identify six task categories: technology, cognitive, interaction, learning, organization, and physical tasks \cite{/content/publication/9789264204027-en}. Participants were asked about their personal and work experiences with these tasks in addition to open-ended questions which were added for completeness, Table \ref{tab:questions_to_cats}.

\begin{table}[h!]
\scriptsize
\centering
\begin{tabular}{p{0.42\columnwidth} p{0.05\columnwidth} |p{0.25\columnwidth}  p{0.05\columnwidth}}
\toprule
\textbf{Age} & \textbf{\#} & \textbf{Gender} &  \textbf{\#} \\
\midrule
18-24 &	22 & Woman & 115 \\
25-34 &	47 & Man & 108 \\ 
35-44 &	43 & Non-binary & 6 \\
45-54 &	40 & No response & 1 \\
55-64 &	60 \\
65 years or older &	18 \\
\midrule
\textbf{Ethnicity} & \textbf{\%} & \textbf{Income} & \textbf{\#} \\
\midrule
American Indian/Alaska Native &	2\% & Under \$30k & 38 \\
Asian or Asian America & 7\% & \$30-\$40k &	30 \\
Black or African American & 12\% & \$40 - \$50k &	13 \\
Hispanic or Latino/a & 10\% & \$50 -\$60k &	22 \\
Native Hawaiʻian/Pacific Islander & <1\% & \$60-\$70k & 17 \\
White or European & 67\% & \$70-\$80k & 20 \\
Other & 1\% & \$80-\$90k & 10 \\
No Response & <1\% & \$90-\$100k & 16 \\
& & Over \$100k & 58 \\
  &  & No response & 5 \\
\midrule
\textbf{Area of Living} & \textbf{\#}  & \textbf{Highest Education} & \textbf{\#} \\
\midrule
Rural & 41 & High school &	38  \\
Suburban & 128 & Some college & 38  \\
Urban & 61 & 2 year degree & 18\\
&  & 4 year degree  & 85 \\
 & & Professional degree & 37  \\
&  &Doctorate & 12  \\
  &  & No response & 1 \\
\bottomrule
\end{tabular}
\caption{Online survey demographics. We had a total of 136 non-adopters (38 who have have never used chat models, 98 who used chat models every few months) and 94 adopters (who used chat models daily). Race demographics are reported in percentages due to multi-racial participants.}
\label{tab:survey_participant_demographics}
\vspace{-2em}
\end{table}

In total, we interviewed 23 participants (Table \ref{tab:participant_demographics}) with limited to no experience with chat models within the United States using snowball sampling via targeted emails \cite{parker2019snowball}, starting with community members from the authors (e.g., in Delaware, Oklahoma, Georgia, California).\footnote{Our seed participants are biased towards author communities, however, the team is highly diverse and our participants also reflect this. Author positionalities [redacted]} We then used a bottom-up approach rooted in grounded theory to qualitatively code the interviews \cite{corbin1990grounded, strauss1997grounded, charmaz2014constructing}, yielding 21 unique needs which we refined, categorized, Table \ref{tab:user_tasks_table}. 

\begin{figure*}
    \centering
    \includegraphics[width=\linewidth]{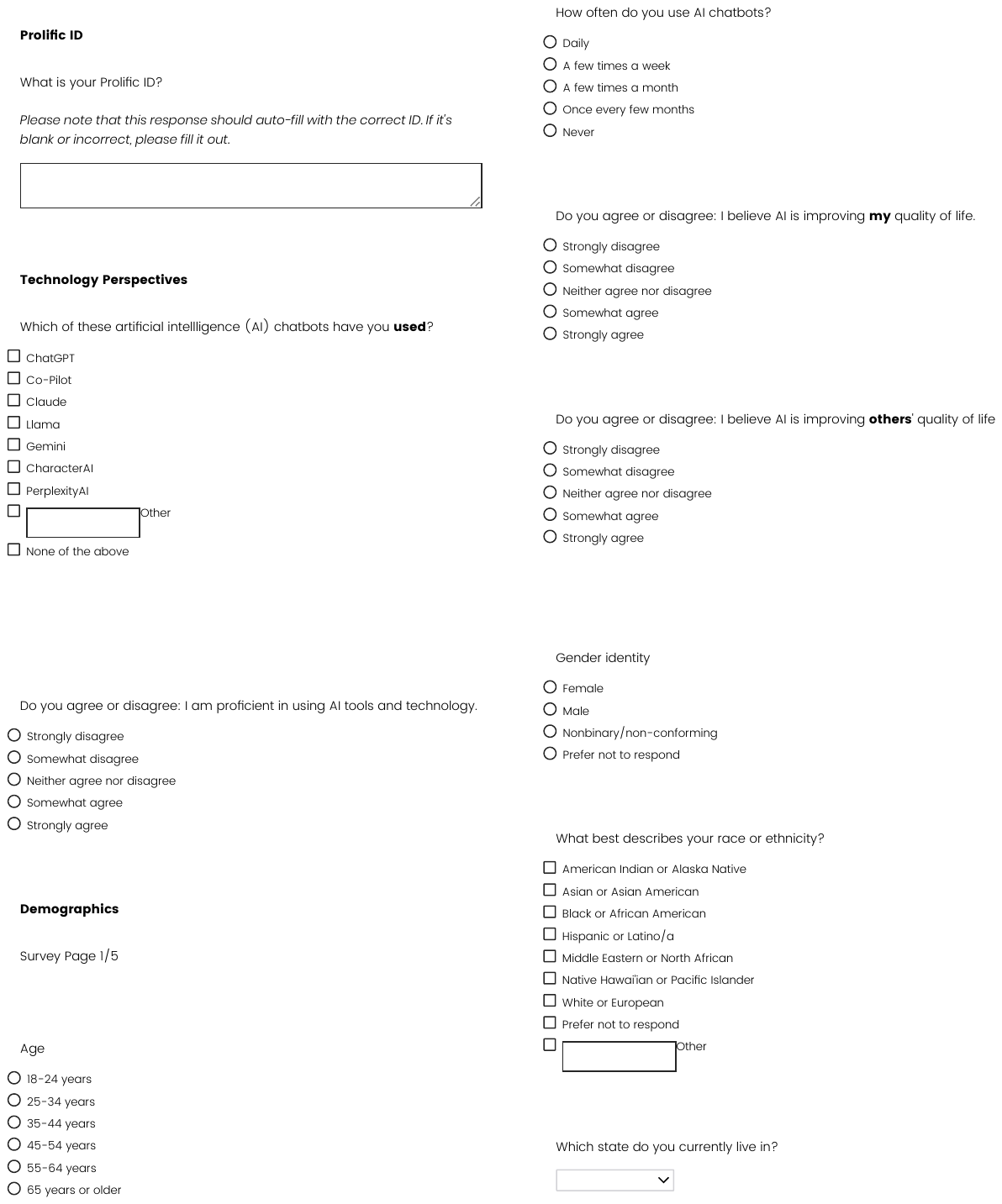}
    \caption{Online survey as presented by Qualtrics}
    \label{fig:online_survey}
\end{figure*}

\paragraph{Online Survey Design}
Informed by our user interviews and literature review, we construct an online survey to understand at a large-scale, how adopters and non-adopters might perceive the importance and painfulness of tasks. We defined non-adopters as participants who never use chat models or use chat models every few months, and adopters are defined as participants who use chat models daily. Our survey included basic demographic questions and questions about task encounter frequency, task importance, and task painfulness, Figure \ref{fig:online_survey}.

Our initial list of tasks was again derived from \textit{The Survey of Adult Skills} with modifications, as shown in Table \ref{tab:questions_to_cats_online_survey}. Creative and caregiving asks were added due to their centrality in the literature review and user interviews, respectively. We asked two questions about task importance and painfulness: participants were asked to select the three most important and painful tasks to their work and personal lives (unordered).

We used Prolific's representative sample recruiting feature to recruit survey participants representative of the U.S. census based on gender, age, and ethnicity. We aimed to recruit a balanced sample of adopters and non-adopters, ultimately having 136 non-adopters and 94 adopters (Table \ref{tab:survey_participant_demographics}).

\begin{figure*}
    \centering
    \includegraphics[width=\linewidth]{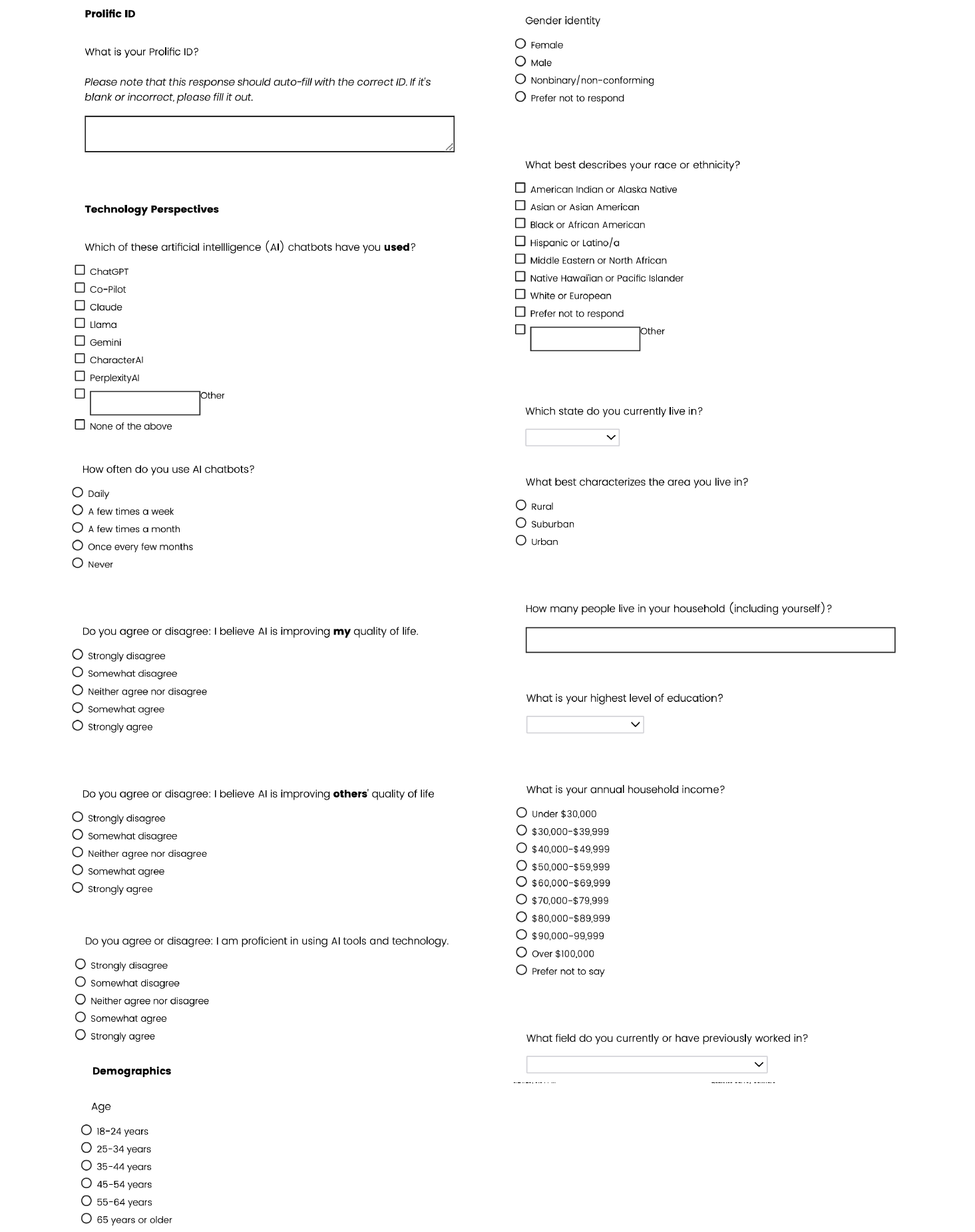}
    \caption{Online Prolific experiment survey as presented by Qualtrics}
    \label{fig:prolific_experiment}
\end{figure*}

\paragraph{Prolific Experiment Design}
\label{sec: prolific_experiment}
We constructed an additional online survey to understand the usage of chat models by Prolific annotators. This survey was IRB approved, and all participants were paid at least \$15 USD an hour. We used Prolific's standard sampling without additional filtering to get an unbiased understanding of the chat model usage by Prolific annotators. This survey included basic demographic questions and questions about chat model usage. In total, we recruited 500 participants. See \ref{fig:prolific_experiment} for survey questions.

\paragraph{Literature Review}
The authors then performed a literature review of the Association of Computational Linguistics (ACL) anthology to identify works that could potentially meet the needs expressed by participants. Although there are extensive works in broader research communities on how various LLMs can be \textit{adapted} to meet a broader set of needs, we situate our literature review in the ACL literature, rather than literature in more applied settings, to review LLM development at its conceptualization, rather than its contextualization. We then mapped non-adopters themes to this literature, clustering tasks into major themes based on how well they address non-adopter needs, resulting in broad categorizations of \textbf{missing} and \textbf{misaligned} tasks.

\paragraph{Pilot Studies: The Difficulties of User Studies}
Prior to formal interviews, we conducted user studies with OpenAI's ChatGPT interface ($n=15$) to build intuition of user needs and to iteratively develop our study protocol. A key insight from these pilots is that \textit{it is very difficult for chat model non-adopters to imagine chat model use cases}. Although demonstrations of how chat models can canonically be used may inspire a new user (e.g., code generation and trivia question answering), it would also bias the user towards thinking that those are the \textit{correct} uses. Instead, we ask participants: What are the greatest pain points in your day-to-day life? 

\paragraph{ChatGPT Demonstrations}
The interviews included a short demonstration of OpenAI's ChatGPT, where the authors created screen-recordings of ChatGPT attempting these tasks that may be of interest to non-adopters (e.g., brainstorming outing activities with an older parent, writing absence notes to school, producing grocery lists with food restrictions). Interviewers also showcased would ask for the participants' reactions, brainstorm other use cases, and debrief on the safe usage of chat models. 

\paragraph{Participant Recruitment and IRB}
For the online survey, we designed the survey on Qualtrics, which took about 5 minutes to complete. Participants were recruited via Prolific and used Prolific's representative sample recruiting feature, which allowed us to recruit participants which were representative of the simplified U.S. census based on gender, age, and ethnicity. The additional Prolific experiment was also designed on Qualtrics and took 1 minute to complete. Participants were recruited via Prolific using the standard sampling recruiting feature, which allowed us to recruit participants who are representative of the broader population on Prolific.
 
\begin{figure*}
    \centering
    \includegraphics[width=\linewidth]{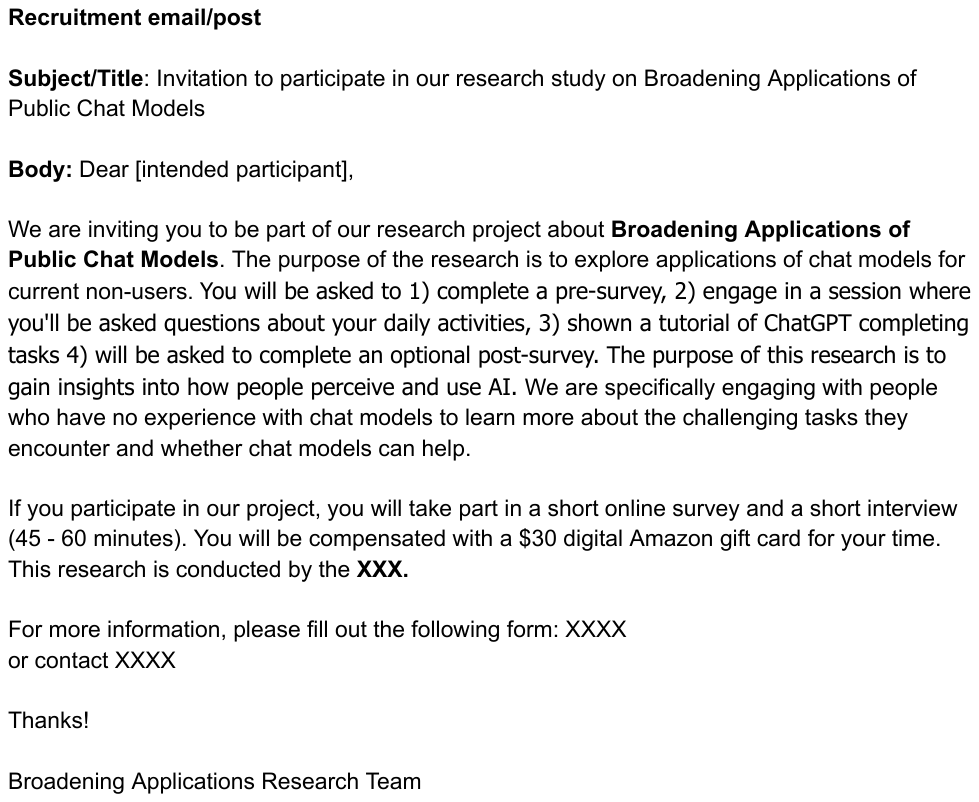}
    \caption{Sample Recruitment Email}
    \label{fig:recruitment_email}
\end{figure*}

\begin{figure*}
    \centering
    \includegraphics[width=\linewidth]{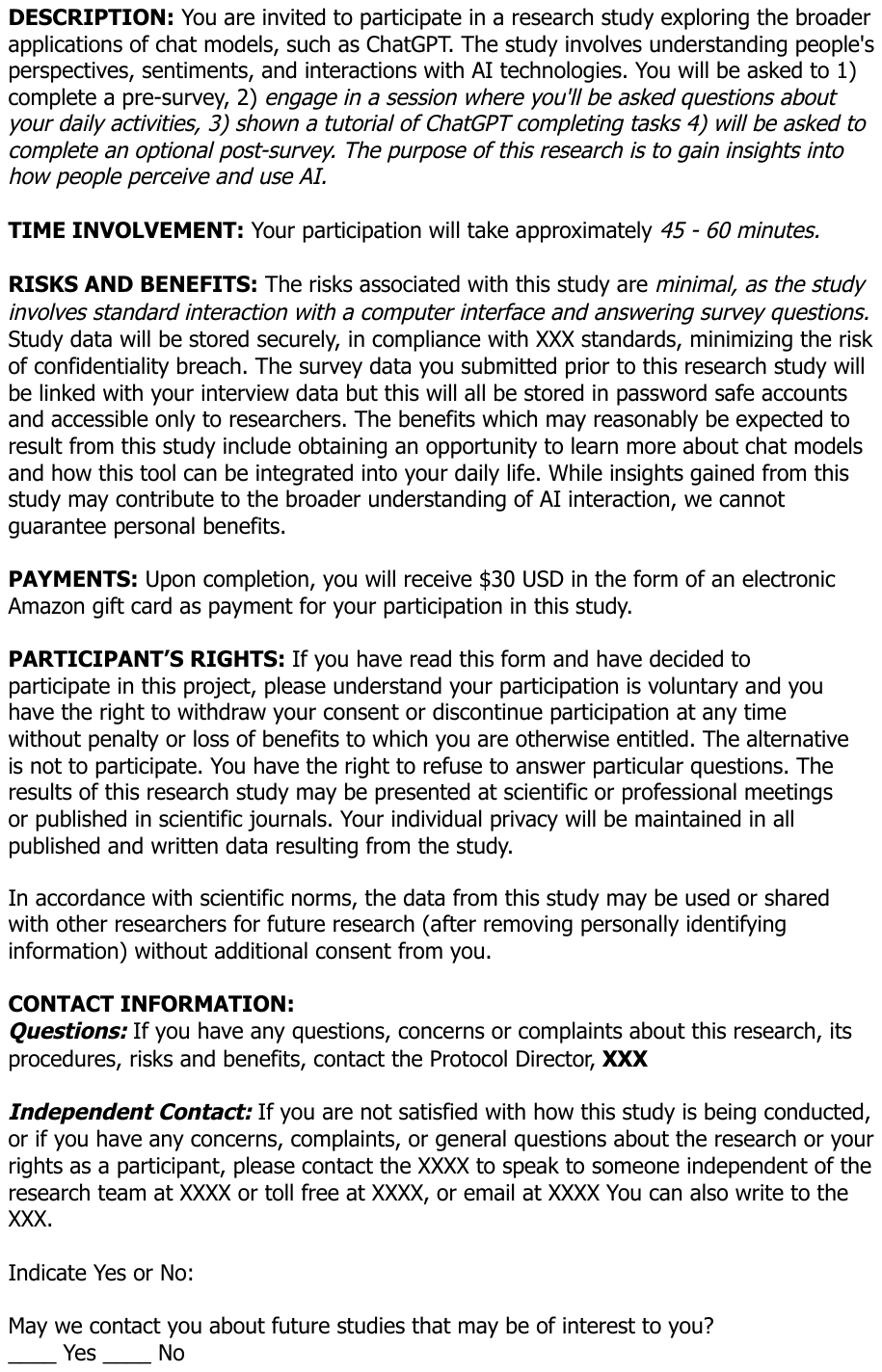}
    \caption{Redacted Consent Form}
    \label{fig:consent_form}
\end{figure*}

\begin{figure*}
    \centering
    \includegraphics[width=\linewidth]{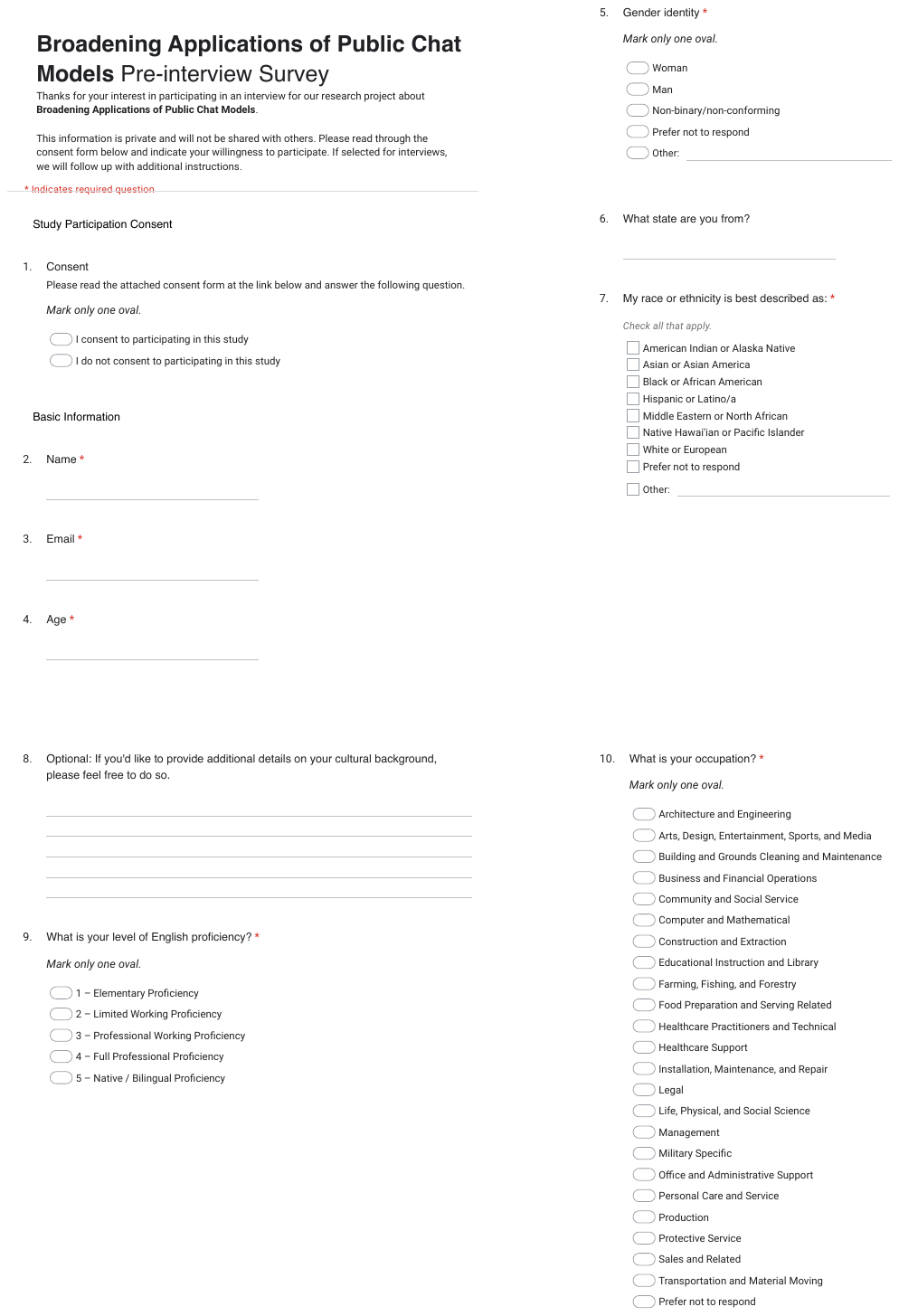}
    \caption{Pre-interview Screener Survey}
    \label{fig:screener_1}
\end{figure*}

\begin{figure*}
    \centering
    \includegraphics[width=\linewidth]{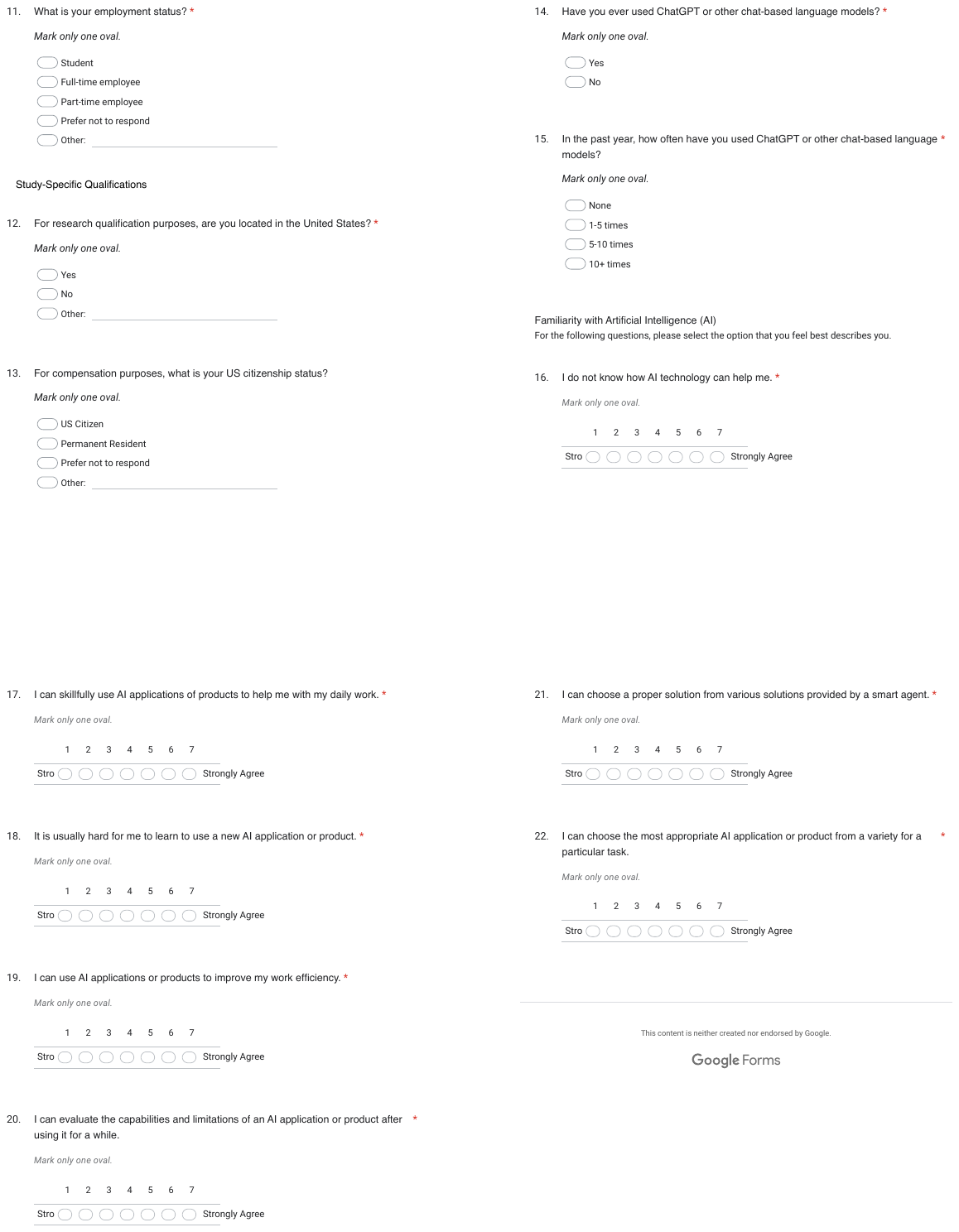}
    \caption{Pre-interview Screener Survey (Part 2)}
    \label{fig:screener_2}
\end{figure*}

For the user interviews, we recruited 23 participants with limited to no experience with chat models within the United States using snowball sampling via targeted emails \cite{parker2019snowball}, starting with community members from the authors (e.g., in Delaware, Oklahoma, Georgia, California).\footnote{Our seed participants are biased towards author communities, however, the team is highly diverse and our participants also reflect this. See author positionalities in Appendix X (redacted for submission).}. We used a pre-interview screener to collect background and demographic information and excluded those who have used chat models over 10 times and automatically included all participants who had never used chat models. Each interviewee received a \$30 USD gift card for a 45-60 minute video-conference interview conducted via Zoom. In total, we interviewed 23 participants from 9 different states, ranging from 19 to 67 years old (Table \ref{tab:participant_demographics}).\footnote{We contacted 15 men and 40 women, but far fewer men were eligible for the study given their prior interactions with chat models, consistent with known skews in LLM usage.} Our participant pool size is considered on par with similar qualitative user studies \cite{smith2024recommend, zhou2022deconstructing, kwon2024unveiling, taranukhin2024empowering, expectation_copilot} which respectively had 13, 18, 12, 15, and 24 participants. This study was IRB approved, and we obtained informed consent from all participants (see Figures \ref{fig:recruitment_email}, \ref{fig:consent_form}, 
\ref{fig:screener_1}, \ref{fig:screener_2} for details).

\paragraph{Qualitative Coding of Themes}
We then used a bottom-up approach rooted in grounded theory to qualitatively code the interviews \cite{corbin1990grounded, strauss1997grounded, charmaz2014constructing}. We iteratively coded and thematically sorted interview excerpts by looking for relations with or among already assigned codes. We first distributed the interview transcripts across all authors for open coding, debriefed in small groups to identify specific needs, and lastly came together to finalize large thematic clusters. Our qualitative coding processes yielded 21 unique needs, which we refined, categorized, and mapped back to our original taxonomy, prioritizing and uplifting needs that were most frequently voiced. Table \ref{tab:user_tasks_table} presents a high-level summary of the concerns voiced by participants. 

\section{Additional Related Work}
\label{sec:more_related_work}
Our work builds on a body of research aiming to make AI and NLP technologies more inclusive \cite{burnett_gender_hci, bardzell2011towards, fiesler_a03, adherng_strengers, blodgett2021sociolinguistically, koenecke2020racial, mun2024particip, zhou2022deconstructing} by adopting participatory  \cite[such as][]{birhane2022power, suresh2024participation, cooper2024constraining, zhi2024beyond, bogiatzis2024beyond, ovadya2024toward} or pluralistic approaches
\cite[such as][]{costanza2020design, birhane2022values, klein2024data, jain2024algorithmic}. Barriers to equity in LLMs include model biases \cite{bolukbasi2016man, schiebinger2014scientific, caliskan2017semantics, sheng2019woman,sap2019risk, blodgett2017racial, zhou2022richer, santurkar2023whose} and performance disparities across languages and ways of speaking \cite{blodgett2016demographic, jurgens-etal-2017-incorporating, koenecke2020racial, ogunremi2023decolonizing},  see \S \ref{sec:more_related_work} for more.

Closest to our work are recent efforts to ground domain-specific applications of LLMs in users' needs and barriers. \citeauthor{kwon_unveiling} identified similar needs around non-trivial information retrieval (e.g., finance and health) and unpaid labor from those with limited ChatGPT experience. \citeauthor{ma2024evaluating} calls for interventions on LLMs at a more foundational level based on evaluations with LGBTQ+ individuals. \citeauthor{mun2024particip} similarly finds that applications for personal life and society outweigh tasks currently focused on in AI development.
HCI research has a long-standing tradition of integrating the distinct needs of various communities into technology design \cite{muller1993participatory, schuler1993participatory, friedman1996value}, addressing aspects such as gender \cite{bardzell2010feminist, burnett_gender_hci, fiesler_a03, adherng_strengers}, race \cite{ogbonnaya2020critical} and \cite{wobbrock2018ability, bayor2021toward}. More recently, we see a rise in engagement in the design of AI systems with relevant stakeholders through the process of co-design with unique user groups such as designing with older adults \cite{sakaguchi2021co}, those with dementia \cite{hsu2023co}, bus drivers \cite{akridge2024bus}, gynecologists \cite{schor2024meaningful}, and rural migrant women \cite{zhao2024design} etc. 

In the fairness community, prior work has called for broadening the participation of LLM design \cite{suresh2024participation, cooper2024constraining, zhi2024beyond}, incorporating pluralist views in AI systems \cite{klein2024data, jain2024algorithmic}, and evaluating the unique risks of technology engagement from minoritized users \cite{grimme2024my, cao2024deleted}. 

Specific to participation in the design of LLMs, HCI works have focused on designing inclusive UX experiences for AI systems~\cite{liao_ux_design}, participatory design with everyday users on LLM hallucination identification~\cite{participatory_design_hallucination}, collaborating with domain experts on policymaking for LLM alignment \cite{feng2024policyprototypingllmspluralistic}, studying how AI could help in responding to harmful online content \cite{mun2024counterspeakers}, and proposing how lay users can surface problematic machine-generated outputs through day-to-day interactions \cite{shen2021everyday}.

\begin{table*}
\scriptsize
\centering
\begin{tabular}{lr|r}
\toprule
\textbf{Task}&\textbf{Given Examples}&\textbf{Original OECD Task Name} \\
\midrule
Technology Navigation&uploading tax documents, scheduling online appointments&Technology - ICT skills \\
Obtain and Understand Information &understanding information related to health or politics&Cognitive Skills - Problem solving \\
Calculation Tasks&budgeting, measurement conversions&Cognitive Skills - Numeracy \\
Writing \& Reading &writing emails, reading documents&Cognitive Skills - Reading \& Writing \\
Learning New Skills&recipes, languages, hobbies&Learning - learning \\
Cooperating, Coordinating, \& Negotiating &shared expenses, planning a family trip&Interaction - co-operation \& influencing \\
Communication Tasks&giving instructions, explaining something&Interaction - co-operation \& influencing \\
Physical Tasks &cleaning, cooking, yard work&Physical - Physical requirement \\
Caring for Others &emotional support, caring for children, sick person, pet&Added - Finding from User Interviews \\
Creative Tasks &designing, imagining, crafting&Added - Prevalence in NLP Literature \\
\bottomrule
\end{tabular}
\caption{List of tasks that survey participants were asked to about in terms of task frequency, importance, and painfulness.}
\label{tab:questions_to_cats_online_survey}
\end{table*}

\begin{table*}[]
\scriptsize
\centering
\begin{tabular}{lr}
\toprule
 \textbf{Question} & \textbf{Task Category} \\
\midrule
\begin{tabular}[c]{@{}l@{}}Q1: In about 3 to 5 sentences, tell me about\\yourself.\end{tabular} & Intro Question \\
\midrule
\begin{tabular}[c]{@{}l@{}}Q2: What are some tasks that are tedious\\to complete that you think take longer\\than they should?\end{tabular} & \begin{tabular}[c]{@{}r@{}}Technology, Cognitive, \\ Learning, Interaction, \\Organization, Physical\end{tabular} \\
\midrule
\begin{tabular}[c]{@{}l@{}}Q3: What are some tasks that are chall-\\enging for you to complete?\end{tabular} & \begin{tabular}[c]{@{}r@{}}Technology, Cognitive, \\  Learning, Interaction,\\Organization, Physical\end{tabular} \\
\midrule
\begin{tabular}[c]{@{}l@{}}Q4: What are some topics that you have\\difficulty obtaining high-quality inform-\\ation for?\end{tabular}  & \begin{tabular}[c]{@{}r@{}}Technology, Cognitive, \\ Learning \end{tabular} \\
\midrule
\begin{tabular}[c]{@{}l@{}}Q5: What are a few things that you wish\\were better explained to you?\end{tabular} & \begin{tabular}[c]{@{}r@{}}Technology, Cognitive, \\ Learning \end{tabular} \\
\midrule
\begin{tabular}[c]{@{}l@{}}Q8: What are some tasks you encounter\\that require reading \& writing?\end{tabular} & Cognitive   \\
\midrule
\begin{tabular}[c]{@{}l@{}}Q7: What are some tasks where you work\\with numbers \& need to make calculations?\end{tabular} & Cognitive  \\
\midrule
\begin{tabular}[c]{@{}l@{}}Q8: What are some creative tasks that you \\do as a part of your work or free time?\end{tabular} & Cognitive \\
\midrule
\begin{tabular}[c]{@{}l@{}}Q9: What are some challenging tasks in \\your life that involve interacting w/ others?\end{tabular} & Interaction, Organization \\
\bottomrule
\end{tabular}
\caption{Interview questions and their respective categories}
\label{tab:questions_to_cats}
\end{table*}

\begin{table*}[]
\scriptsize
\centering
\begin{tabular}{p{0.38\columnwidth} p{0.05\columnwidth} |p{0.35\columnwidth} p{0.015\columnwidth}}
\toprule
\textbf{Age} & \textbf{\#} & \textbf{Gender} &  \textbf{\#} \\
\midrule
18-19 &	1 & Woman & 19 \\
20-29 &	4 & Man & 4 \\ 
30-39 &	3 & Non-binary & 0 \\
40-49 &	6 & Prefer not to respond & 0 \\
50-59 &	8 \\
60-69 &	1 \\
\midrule
\textbf{Ethnicity} & \textbf{\%} & \textbf{U.S. Region} & \textbf{\#} \\
\midrule
American Indian/Alaska Native &	9\% & Middle Atlantic (NJ,PA) & 	2 \\
Asian or Asian America & 26\% & East North Central (WI) &	1 \\
Black or African American & 22\% & East South Central (KY) &	1 \\
Hispanic or Latino/a & 9\% & South Atlantic (DE,FL,GA) &	9 \\
Middle Eastern/North African	& 4\% & West South Central (OK)	& 5 \\
White or European & 43\% & Pacific	(CA) & 5 \\
Preferred not to Respond & 4\% & & \\
\midrule
\textbf{Occupation} & \textbf{\#} & \textbf{Chat Model Usage} & \textbf{\#} \\
\midrule
Architecture and Engineering & 2 & None & 	14 \\
Business and Financial & 3 & 1-5 times	& 4 \\
Education and Library	& 3 & 5-10 times	& 4 \\
Health Care &	7 & 10+ times\footnote{The participant confused usage frequency with conversation turns and in fact, had only used chat models a couple of times.} & 1 \\
Sales and Related &	2 & & \\
Other &	6 & & \\

\bottomrule
\end{tabular}
\caption{Participant demographics as self-reported via pre-interview screener. Race demographics are reported in percentages as counts do not add up to 23 due to multi-racial participants.}
\label{tab:participant_demographics}
\end{table*}

\begin{table*}[h!]
\scriptsize
\begin{tabular}{llllcllllll}
\toprule
& \multicolumn{2}{c}{\textbf{TECHNOLOGY}} & \multicolumn{2}{c}{\textbf{COGNITIVE}} & \multicolumn{2}{c}{\textbf{INTERACTION}} & \multicolumn{2}{c}{\textbf{ORGANIZATION}} & \multicolumn{2}{c}{\textbf{PHYSICAL}} \\
\midrule
& \multicolumn{1}{c}{\begin{tabular}[c]{@{}c@{}}Software\\ Interaction\end{tabular}} & \multicolumn{1}{c}{\begin{tabular}[c]{@{}c@{}}Continuous \\ Learning\end{tabular}} & \multicolumn{1}{c}{\begin{tabular}[c]{@{}c@{}}Non-Trivial \\ Information \\ Retrieval\end{tabular}} & \begin{tabular}[c]{@{}c@{}}Non-Needs:\\ Calculations \&\\ Creativity\end{tabular} & \multicolumn{1}{c}{\begin{tabular}[c]{@{}c@{}}Cross-Cultural\\ Communi-\\cation\end{tabular}} & \multicolumn{1}{c}{\begin{tabular}[c]{@{}c@{}}Providing \\ Emotional \\ Support\end{tabular}} & \multicolumn{1}{c}{\begin{tabular}[c]{@{}c@{}}Domestic \\ Work\end{tabular}} & \multicolumn{1}{c}{\begin{tabular}[c]{@{}c@{}}Unpaid \\ Labor\end{tabular}} & \multicolumn{1}{c}{\begin{tabular}[c]{@{}c@{}}Accommo-\\ dations\end{tabular}} & \multicolumn{1}{c}{\begin{tabular}[c]{@{}c@{}}Physical\\ Chores\end{tabular}} \\
\midrule
P1 & \multicolumn{1}{c}{X} & \multicolumn{1}{c}{X} & & X & & & & & \multicolumn{1}{c}{X} & \\
\arrayrulecolor{black!30}\hline
P2 & & \multicolumn{1}{c}{X} & & \multicolumn{1}{l}{} & \multicolumn{1}{c}{X} & & & & & \multicolumn{1}{c}{X} \\
\arrayrulecolor{black!30}\hline
P3 & & & & X & \multicolumn{1}{c}{X} & & & & & \\
\arrayrulecolor{black!30}\hline
P4 & \multicolumn{1}{c}{X} & \multicolumn{1}{c}{X} & & \multicolumn{1}{l}{} & & & & & & \\
\arrayrulecolor{black!30}\hline
P5 & & & \multicolumn{1}{c}{X} & X & \multicolumn{1}{c}{X} & & & & & \\
\arrayrulecolor{black!30}\hline
P6 & & \multicolumn{1}{c}{X} & & X & & & & & \multicolumn{1}{c}{X} & \\
\arrayrulecolor{black!30}\hline
P7 & & \multicolumn{1}{c}{X} & & X & & \multicolumn{1}{c}{X} & & \multicolumn{1}{c}{X} & \multicolumn{1}{c}{X} & \\
\arrayrulecolor{black!30}\hline
P8 & \multicolumn{1}{c}{X} & \multicolumn{1}{c}{X} & & \multicolumn{1}{l}{} & & & \multicolumn{1}{c}{X} & & & \\
\arrayrulecolor{black!30}\hline
P9 & & \multicolumn{1}{c}{X} & & X & & & & & & \\
\arrayrulecolor{black!30}\hline
P10 & & & \multicolumn{1}{c}{X} & X & \multicolumn{1}{c}{X} & \multicolumn{1}{c}{X} & \multicolumn{1}{c}{X} & \multicolumn{1}{c}{X} & & \\
\arrayrulecolor{black!30}\hline
P11 & \multicolumn{1}{c}{X} & \multicolumn{1}{c}{X} & & X & & & & & & \\
\arrayrulecolor{black!30}\hline
P12 & & & \multicolumn{1}{c}{X} & X & & & \multicolumn{1}{c}{X} & & & \\
\arrayrulecolor{black!30}\hline
P13 & \multicolumn{1}{c}{X} & & & \multicolumn{1}{l}{} & \multicolumn{1}{c}{X} & & & & & \multicolumn{1}{c}{X} \\
\arrayrulecolor{black!30}\hline
P14 & & & & X & & \multicolumn{1}{c}{X} & \multicolumn{1}{c}{X} & \multicolumn{1}{c}{X} & & \\
\arrayrulecolor{black!30}\hline
P15 & & & & X & & & \multicolumn{1}{c}{X} & & & \multicolumn{1}{c}{X} \\
\arrayrulecolor{black!30}\hline
P16 & \multicolumn{1}{c}{X} & & \multicolumn{1}{c}{X} & \multicolumn{1}{l}{} & & \multicolumn{1}{c}{X} & \multicolumn{1}{c}{X} & \multicolumn{1}{c}{X} & \multicolumn{1}{c}{X} & \\
\arrayrulecolor{black!30}\hline
P17 & & & & \multicolumn{1}{l}{} & & \multicolumn{1}{c}{X} & \multicolumn{1}{c}{X} & \multicolumn{1}{c}{X} & \multicolumn{1}{c}{X} & \\
\arrayrulecolor{black!30}\hline
P18 & & & & \multicolumn{1}{l}{} & & & & \multicolumn{1}{c}{X} & & \\
\arrayrulecolor{black!30}\hline
P19 & & & \multicolumn{1}{c}{X} & X & & & & & & \multicolumn{1}{c}{X} \\
\arrayrulecolor{black!30}\hline
P20 & \multicolumn{1}{c}{X} & \multicolumn{1}{c}{X} & & X & \multicolumn{1}{c}{X} & \multicolumn{1}{c}{X} & \multicolumn{1}{c}{X} & \multicolumn{1}{c}{X} & & \\
\arrayrulecolor{black!30}\hline
P21 & & & & X & & & & \multicolumn{1}{c}{X} & & \multicolumn{1}{c}{X} \\
\arrayrulecolor{black!30}\hline
P22 & & & & X & \multicolumn{1}{c}{X} & & & \multicolumn{1}{c}{X} & & \\
\arrayrulecolor{black!30}\hline
P23 & & & & X & & \multicolumn{1}{c}{X} & \multicolumn{1}{c}{X} & \multicolumn{1}{c}{X} & & \\
\arrayrulecolor{black} \midrule
& \multicolumn{1}{c}{30\%} & \multicolumn{1}{c}{39\%} & \multicolumn{1}{c}{22\%} & 70\% & \multicolumn{1}{c}{30\%} & \multicolumn{1}{c}{30\%} & \multicolumn{1}{c}{39\%} & \multicolumn{1}{c}{43\%} & \multicolumn{1}{c}{22\%} & \multicolumn{1}{c}{22\%} \\
\bottomrule
\end{tabular}
\caption{Taxonomy of user needs and occurrences for each participant.
\textbf{Software interaction}: interacting with existing software systems
\textbf{Continuous learning}: constantly learning new technologies
\textbf{Non-Trivial Information Retrieval}: finding or making sense of information unanswerable by search queries alone
\textbf{Cross Cultural Communication}: digital and real life communication across cultures and generations
\textbf{Providing Emotional Support}: supporting others
\textbf{Domestic Work}: domestic duties, including the mental load
\textbf{Unpaid labor}: historically unpaid labor e.g., organizing, coordinating, planning
\textbf{Accommodations}: seeking out and learning about physical accommodations
\textbf{Physical Chores}: routine physical chores
\textbf{Non-needs:} pain-free cognitive tasks like calculations and creativity
}

\label{tab:user_tasks_table}
\end{table*}

\begin{figure*}
    \centering
    \includegraphics[width=\linewidth]{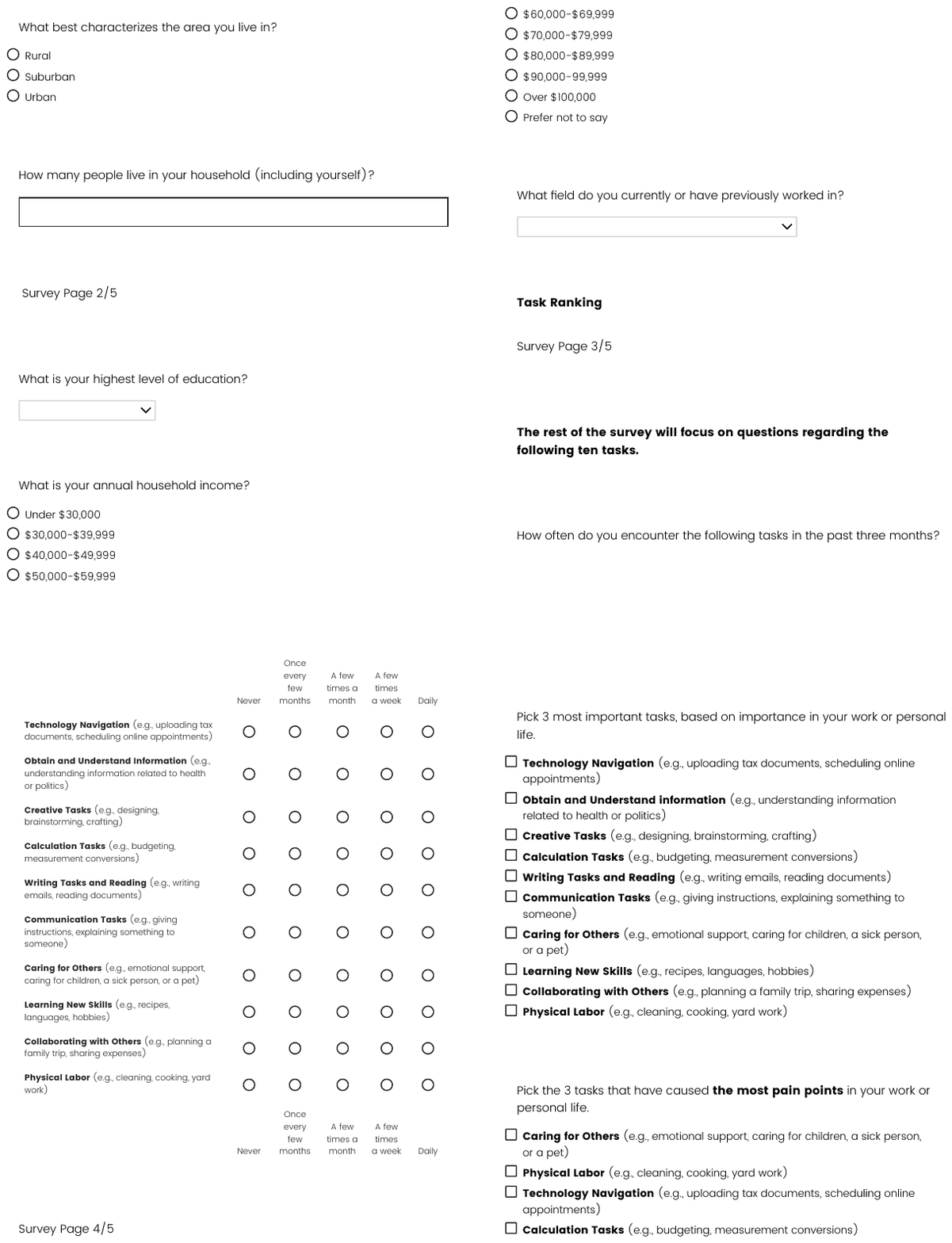}
    \caption{Online survey as presented by Qualtrics}
    \label{fig:online_survey2}
\end{figure*}

\begin{figure*}
    \centering
    \includegraphics[width=\linewidth]{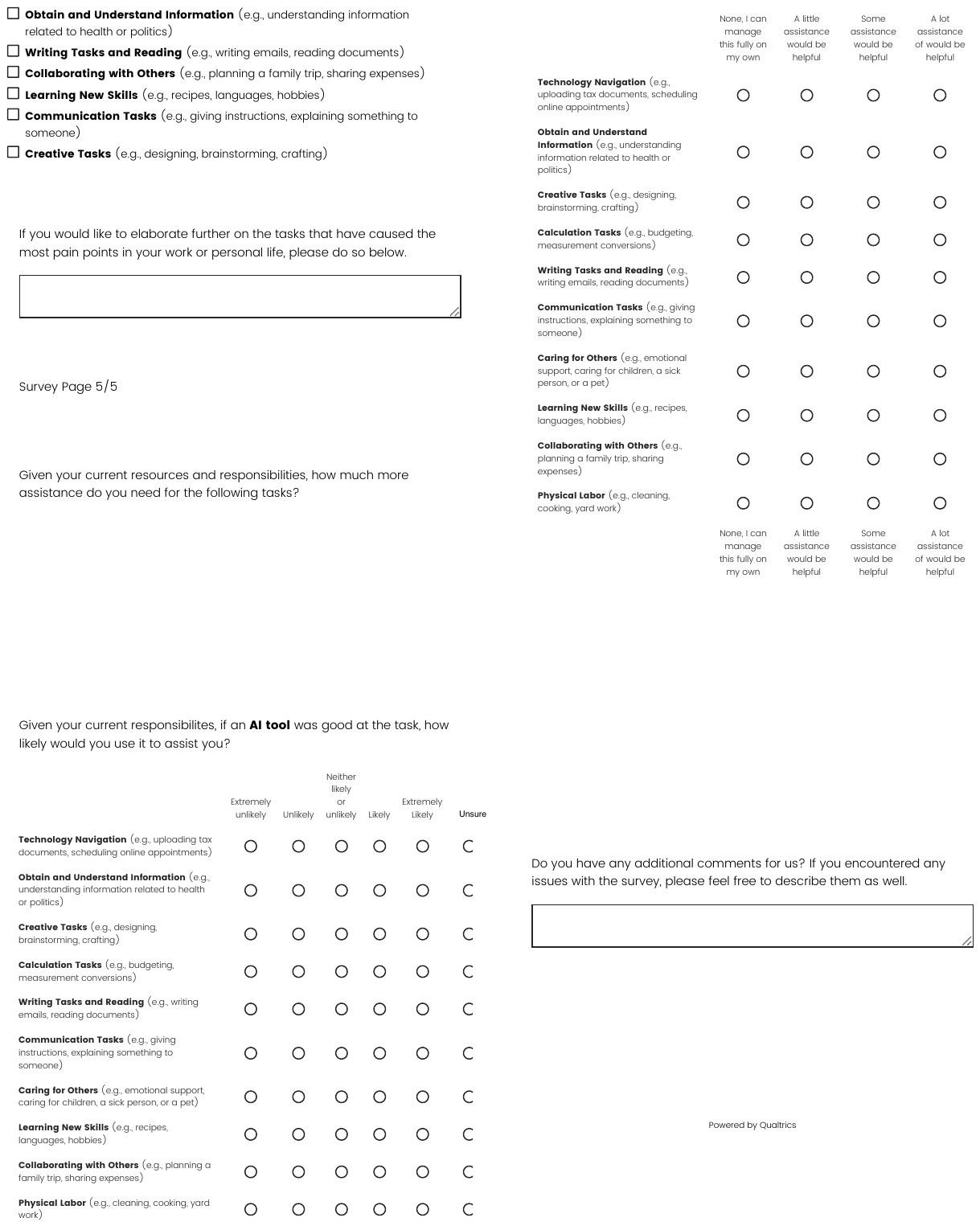}
    \caption{Online survey as presented by Qualtrics}
    \label{fig:online_survey3}
\end{figure*}

\begin{figure*}
    \centering
    \includegraphics[width=\linewidth]{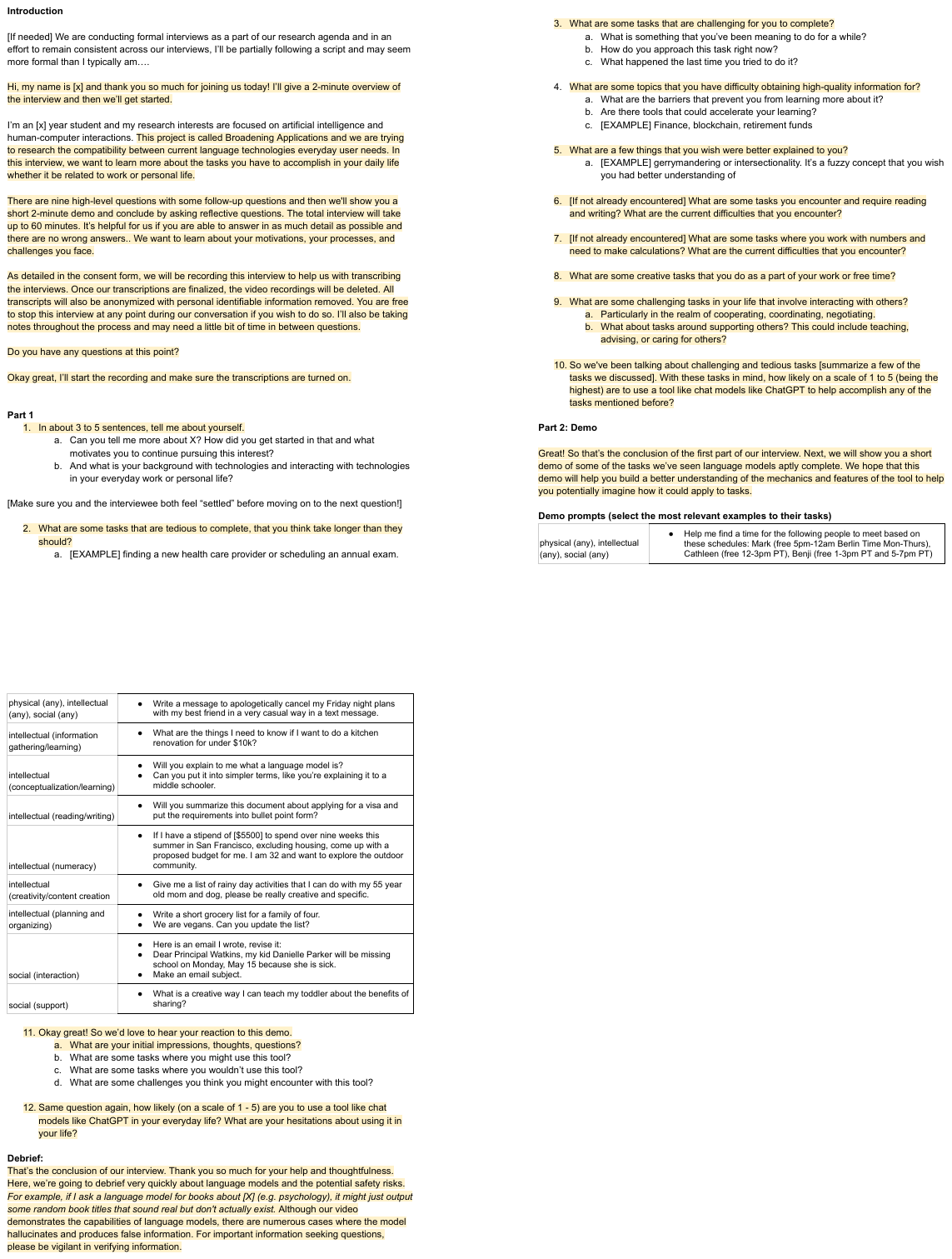}
    \caption{Interview script read by interviewers. Highlighted parts are asked verbatim.}
    \label{fig:interview_script}
\end{figure*}

\begin{figure}
    \centering
    \includegraphics[width=\linewidth]{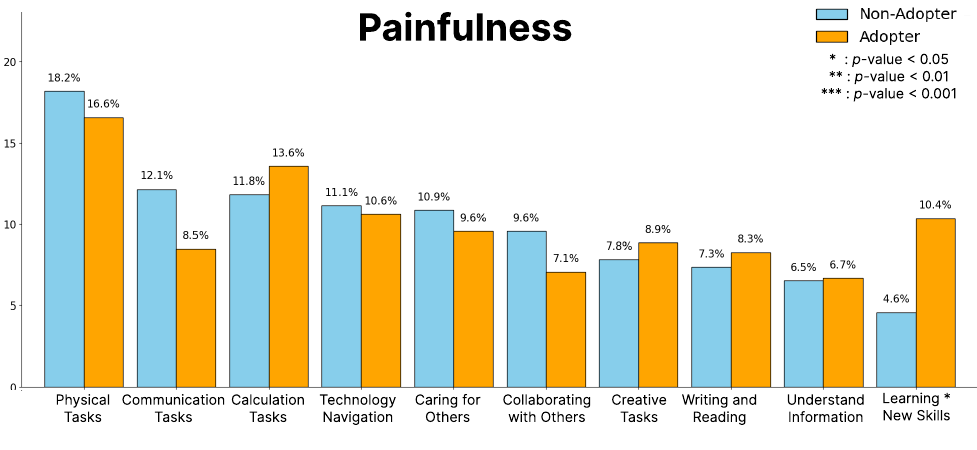}
    \caption{This figure shows task painfulness rankings between LLM adopters and non-adopters. Adopters regarded learning new skills as more painful compared to non-adopters. However, many tasks don't have a significant difference between adopters and non-adopters.}
    \label{fig:painfulness}
\end{figure}

\begin{figure}
    \centering
    \includegraphics[width=\linewidth]{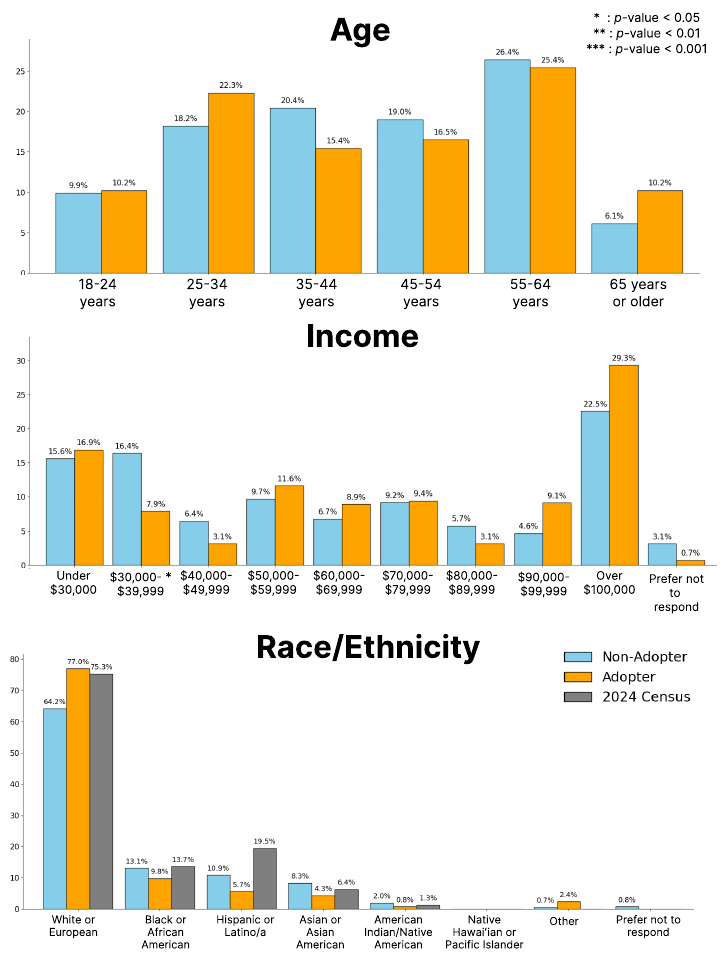}
    \caption{This chart visualizes age, income, and race/ethnicity across non-adopter and adopters. Non-adopters and adopters were similar across age groups. Non-adopters tended to earn under \$70,000 annually, whereas adopters tended to be 18-34 years old.}
    \label{fig:demographics_appendix}
\end{figure}

\begin{figure}
    \centering
    \includegraphics[width=\linewidth]{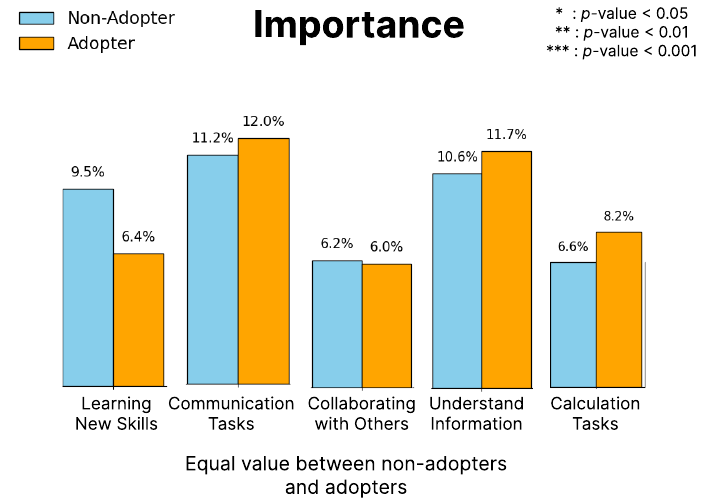}
    \caption{This figure shows task importance rankings between chat model adopters and non-adopters. These tasks show no significant difference in prioritization between adopters and non-adopters.}
    \label{fig:importance2}
\end{figure}

\end{document}